\documentclass{revtex4}
\usepackage{sidecap}
\usepackage{amsmath}
\usepackage{graphicx}
\begin{document}
\newcommand{\be}{\begin{equation}}
\newcommand{\ee}{\end{equation}}
\newcommand{\beq}{\begin{eqnarray}}
\newcommand{\eeq}{\end{eqnarray}}
\markboth{B.Haskell \& A.Melatos}
{Models of Pulsar Glitches}

%
%

\title{Models of Pulsar Glitches }

\author{B. Haskell}
\address{School of Physics, University of Melbourne, Parkville, VIC 3010, Australia}
\author{A. Melatos}
\address{School of Physics, University of Melbourne, Parkville, VIC 3010, Australia}



\begin{abstract}
Radio pulsars provide us with some of the most stable clocks in the universe. Nevertheless several pulsars exhibit sudden spin-up events, known as glitches. More than forty years after their first discovery, the exact origin of these phenomena is still open to debate. It is generally thought that they an observational manifestation of a superfluid component in the stellar interior and provide an insight into the dynamics of matter at extreme densities. In recent years there have been several advances on both the theoretical and observational side, that have provided significant steps forward in our understanding of neutron star interior dynamics and possible glitch mechanisms.  In this article we review the main glitch models that have been proposed and discuss our understanding, in the light of current observations. 
\end{abstract}




\maketitle

Radio pulsars are thought to be rotating magnetised neutron stars (NS). The huge moment of inertia (of the order of $10^{45}$ g cm$^2$) leads to exceptionally stable rotation rates and provides us with some of the most precise clocks in the universe. The best timed millisecond pulsars are stable to a precision which rivals that of atomic clocks \cite{Hobbsclocks}. Nevertheless radio pulsars exhibit several timing irregularities, the most striking of which are the so-called glitches. While most objects are observed to steadily spin-down due to the emission of electromagnetic and, possibly, gravitational waves, many pulsars show sudden increases in their spin, in some cases followed by an increase in their spin-down rate, i.e. glitches. Most pulsars also show slower, long-term, stochastic deviations from a regular spin-down law, that are generally classed as `timing noise' and are not the main focus of this review article.

Soon after the discovery of the first glitches in the Vela \cite{RM69, RD69} and Crab \cite{Crab1, Crab2} pulsars in 1969, several mechanisms were suggested to explain these phenomena. Although some initial suggestions featured external mechanisms, such as plasma explosions in the magnetosphere \cite{pacini} or planets around the pulsar \cite{planets}, the lack of radiative and pulse profile changes associated with these events was taken as evidence for an internal origin. The situation is different for rotating radio transients (RRATs) and magnetars, which are now known to glitch and do exhibit, amongst other peculiar traits, radiative changes associated with these events \cite{Lyne10, Liv11, Welte11, Keith13}. Magnetospheric activity is likely to play a role in these objects, as we discuss below.

There are two main internal glitch mechanisms that have been examined in the literature.
 The first set of models relies on the fact that the outer layers of a neutron star form a crystalline crust that can support stress. As the NS spins down, the liquid core adjusts its shape to the rotation rate, while the solid crust maintains the shape appropriate for the previous, higher, spin rate. This leads to an increasing amount of strain building up in the crust, which is eventually released in the form of a star quake. The quake causes a sudden rearrangement of the moment of inertia and ultimately a glitch \cite{Rud69,Small70}.

A second set of models is based on the prediction that neutrons in the interiors of mature NSs are superfluid \cite{Migdal, BPPsuper}, a prediction that has been supported recently by monitoring the cooling of the young NS in the supernova remnant Cassiopeia A \cite{casa1, casa2}. At the high densities in the stellar interior (which can exceed nuclear saturation density in the core), neutrons find it energetically favourable to pair, thus creating an energy gap below the Fermi surface, that must be bridged if the particles are to interact and take part, for example, in the scattering processes that give rise to viscosity.
Superfluidity thus suppresses the viscosity and lengthens the coupling time-scale between the neutron superfluid and the non-superfluid crust. The long recovery time-scales (of order months) observed after the first Vela glitch were, in fact, immediately considered to be evidence for a weakly coupled superfluid component in the stellar interior \cite{2fluid}.

Another key property of a superfluid is that its flow is irrotational. The system of neutron pairs is similar to a Bose-Einstein condensate, described by a macroscopic wave function, whose momentum is the gradient of a phase. Nevertheless we know that the star rotates rapidly and that interactions with the normal component, however weak, induce rotation in the neutron superfluid. It is well established from the study of superfluid helium that a superfluid rotates by creating an array of quantised vortices that carry the circulation. The angular velocity of an element of fluid is proportional to the number of vortices it contains. For it to spin down, some of them must be removed.
Anderson and Itoh \cite{AI} suggested that interactions between vortices and ions in the NS crust can `pin' the vortices and restrict their outward motion. As long as the vortices are pinned, i.e. stay fixed in position, the superfluid does not spin down, thereby storing angular momentum, which is periodically released in glitches.

The vortex model has become the standard picture for pulsar glitches. This is partially due to the crust-quake model's inability to explain the large and frequent glitches now observed in the Vela pulsar \cite{Flan96}, but mainly due to the success of the vortex creep model developed by Alpar and collaborators \cite{Alpar84a} in explaining post facto the observed post-glitch relaxation of the Vela and other pulsars in terms of general parameters of NS structure. Furthermore recent simulations based on this scenario have successfully described several aspects of Vela giant glitches \cite{Haskell12} and the statistics of the glitching populations as a whole \cite{Lila11, lilaknock}. Nevertheless several issues still need to resolved before the vortex picture attains the status of a self-consistent, predictive, falsifiable theory. The trigger for vortex unpinning is still unknown and may be due to vortex accumulation in strong pinning areas \cite{alpartrigger, pierre}, vortex domino effects \cite{lilaknock}, hydrodynamical instabilities \cite{Prix03,Peraltaglitch,Nilsglitch} or quakes \cite{rudermanpinquake,alparquake1}.
Recent calculations have also showed that Bragg scattering severely limits the mobility of neutrons in the crust, limiting the amount of angular momentum that can be stored in the crust between glitches \cite{Chamel1}. An analysis of the Vela pulsar reveals that in this case it is difficult to accommodate the implied angular momentum exchange during a glitch unless the pulsar has a low mass $\lesssim 1 M_\odot$ or part of the core is involved in the process \cite{Chamel2, crustnot}. The core of the NS is, in fact, thought to be in a type II superconducting state\cite{BPPsuper}. Neutron vortices could then interact with proton flux tubes leading to observable consequences \cite{Linkresponse}.


\section{Neutron star interiors}

\label{neutron}
Neutron stars are one of the most extreme and exciting nuclear physics laboratories in the universe. With a mass slightly above that of the Sun compressed into a radius of $\approx$ 10 km, their interiors can easily exceed nuclear saturation density, allowing us to probe many-body aspects of the strong interaction at MeV energies, which cannot be studied in terrestrial laboratories. The equation of state of matter at such high densities is not known, and several possibilities exist. Recent observations of NS masses slightly above $2 M_{\odot}$ \cite{2M1, 2M2} begin to constrain our understanding of physics in this regime. The general relativistic equations of hydrostatic equilibrium, unlike their Newtonian equivalent, lead to a maximum mass as a function of radius (or equivalently central density) which is equation of state dependent. In figure \ref{f1} we plot the mass-radius relation for several equations of state that allow for a maximum mass above $2 M_\odot$. Softer equations of state (with low maximum mass) are excluded, which constrains the presence of exotica (such as hyperons) in the core\cite{Lat2m}.

Furthermore at such high densities the Fermi energy is large compared to the thermal energy so that, although mature NSs have an internal temperature $\gtrsim 10^7$ K, they behave essentially as cold objects. In most of the star it is energetically favourable for neutrons to be superfluid, while protons in the core are superconducting. In figure \ref{f1} we show the critical transition temperatures for neutron superfluidity and proton superconductivity for a representative sample of models. As can be seen the interior of a mature NS is expected to be mostly superfluid. Note, however, that the density dependence of the pairing gaps is such that there will always be regions close to the transition temperature in which thermal effects cannot be neglected. 

Superfluidity impacts on the star in many ways. It modifies transport coefficients (e.g.\ thermal conductivity, kinematic viscosity) \cite{SFconduct, NilsKostasVisc}, it allows neutrons and protons to flow independently, and the neutron condensate forms an array of quantised vortices (an Abrikosov lattice) which carries the vorticity \cite{BPPsuper}. We discuss the effects of superfluidity in detail in the following sections.

\begin{figure}
\centerline{\includegraphics[width=8cm]{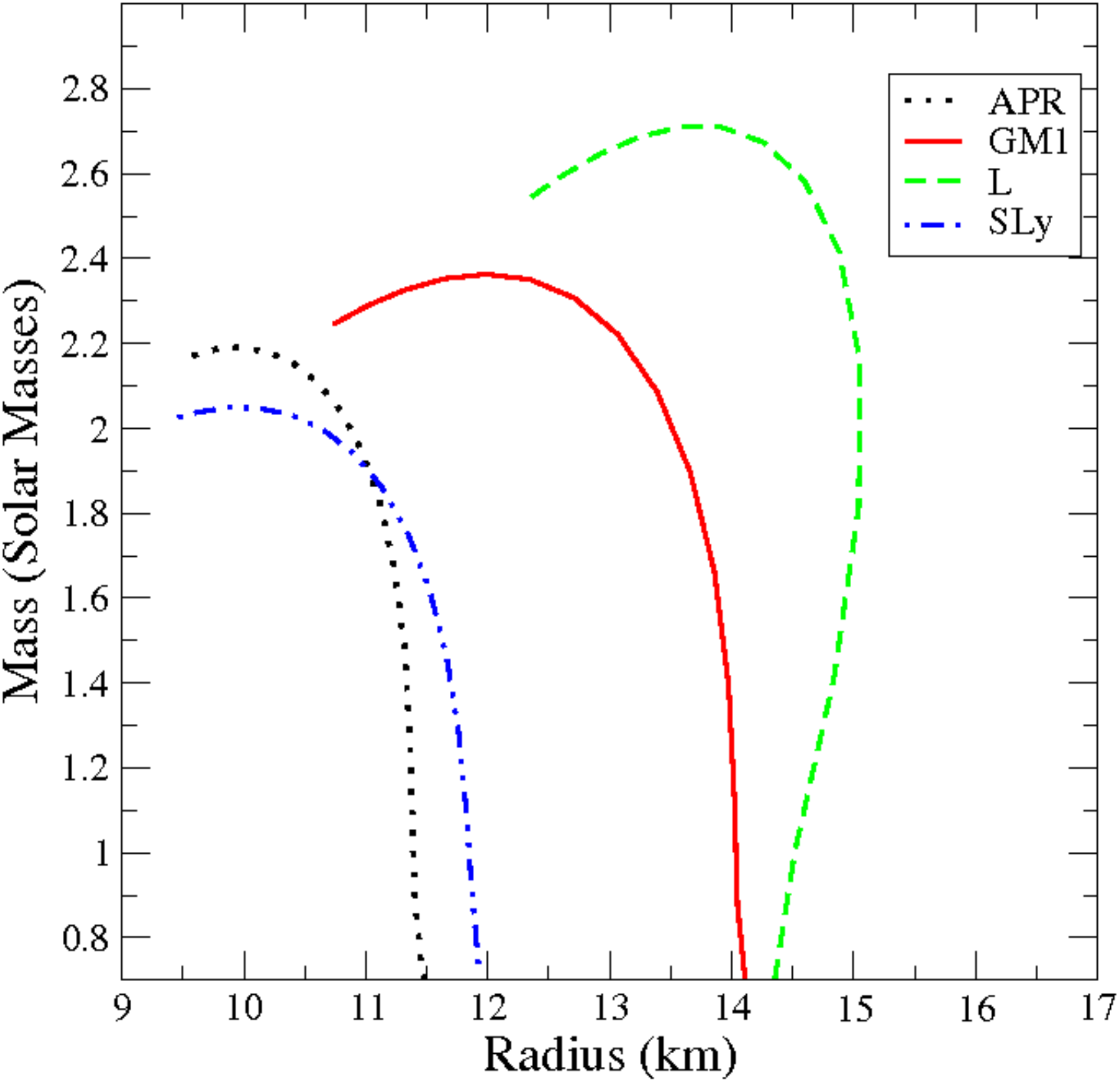}\;\;\;\;\;\;\includegraphics[width=8.2cm]{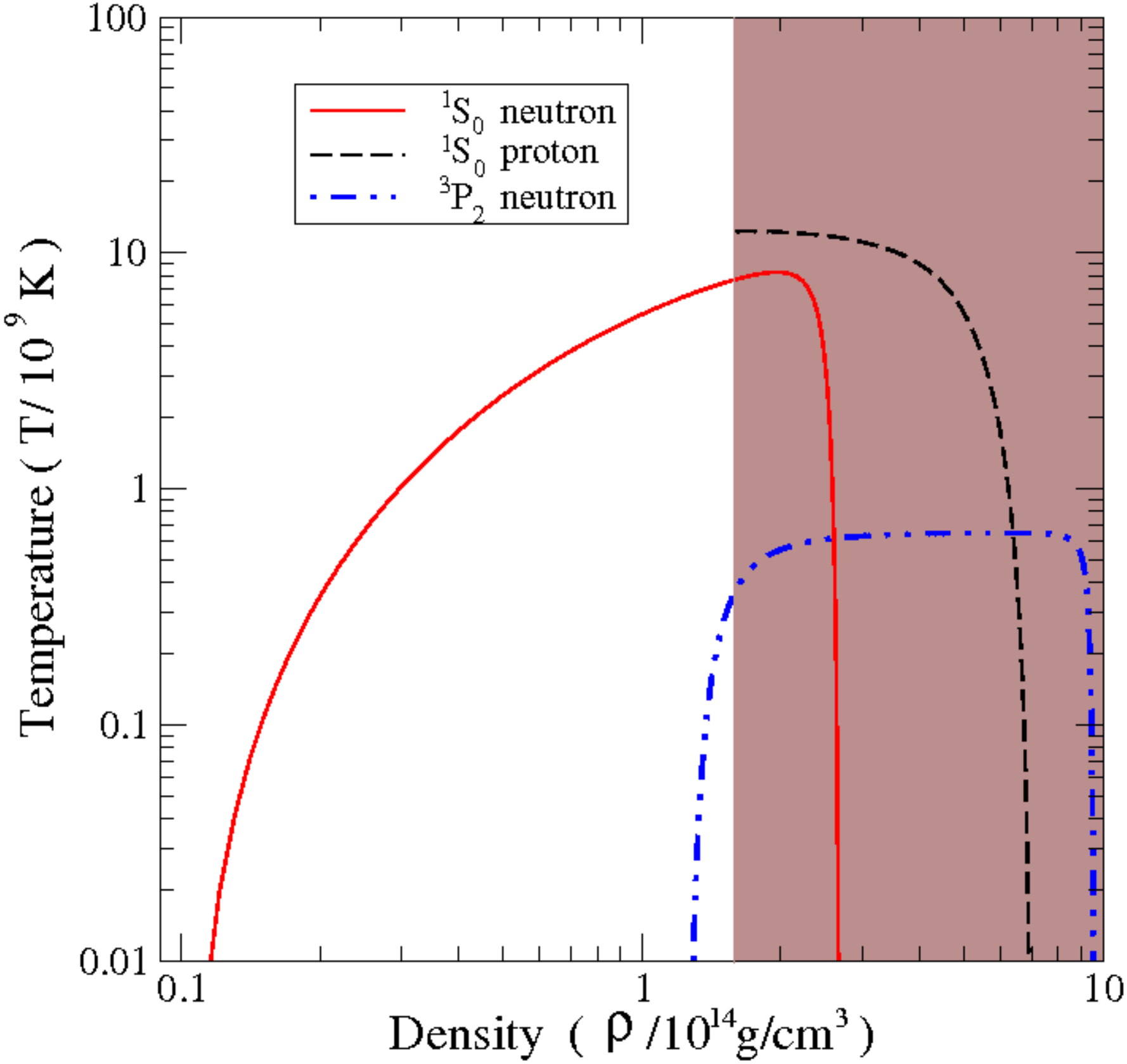}}
\vspace*{8pt}
\caption{LEFT: The mass-radius relation for four representative equations of state: SLy \cite{SLy}, APR \cite{APR}, GM1 \cite{GM1} and L \cite{ABL}. As can be seen the chosen equations of state satisfy the observational constraint of allowing for a maximum mass greater than $2 M_\odot$.
RIGHT: The critical transition temperature for a representative set of gap models, taken from \cite{NilsKostasVisc} (models d,g and l). In the crust one has neutron pairing in the $^1S_0$ channel, with $^3P_2$ pairing possible at higher densities in the core. We plot also the $^1S_0$ gap for protons in the core. The critical temperature for  $^3P_2$ pairing in the core is constrained to be in the region $(5-8) \times 10^8$ K by observations of the cooling of the NS in Cassiopeia A \cite{casa1,casa2}. The shaded region represents the core. \label{f1}}
\end{figure}

\subsection{Layered structure}

The outer regions of the star comprise a solid crust, in which ions are arranged in a crystalline lattice which, above a density of $\approx 5\times 10^{11} \, {\rm g\, cm^{-3}}$, is immersed in a gas of free (superfluid) neutrons. This region thus behaves as an elastic solid and has a finite shear modulus $\mu$, which can be calculated if the equation of state and composition of the crust are known \cite{ogata90}:
\be
\mu=0.1194\left(\frac{4\pi}{3}\right)^{1/3}\left[ \frac{(1-X_\mathrm{n}) n_b }{A} \right]^{4/3} (Ze)^2\;\;\mbox{g/(cm s$^2$)},
\ee
where $X_n$ is the fraction of neutrons outside nuclei, $n_b$ is the baryon number density, $A$ and $Z$ the atomic and proton number, and $e$ the fundamental charge. For typical compositions and densities \cite{douchinhaensel} this gives a shear modulus of the order $\mu\approx 10^{30}$ g cm$^{-1}$ s$^{-2}$.
The maximum elastic stress, $\bar{t}_m$, that the crust can sustain before cracking takes the form
\be
\bar{t}_m\approx \mu \bar{\sigma}_m
\ee
where $\bar{\sigma}_m$ is the breaking strain. By analogy with terrestrial crystals, Smoluchowski and Welch\cite{SW70} suggested that $10^{-5} \lesssim \bar{\sigma}_m \lesssim 10^{-3}$, but recent molecular dynamics simulations \cite{Horowitz} imply $\bar{\sigma}_m\approx 0.1$. The latter simulations also suggest that the system is dominated by pressure and gravity, and the crust does not crack along fault lines, as in terrestrial earthquakes, but rather fails as a whole once the maximum strain is exceeded.

Note that the ions are generally taken to be arranged in a Body Centered Cubic (BCC) lattice, however there is still significant uncertainty on the exact composition and structure of the crust \cite{Steiner}, and not only electrons, but also free neutrons, may partially screen the Coulomb interaction between the nuclear clusters, leading to different, and more inhomogeneous, configurations than a BCC lattice \cite{JonesAm, KP14}. This can have important consequences for the interaction between vortices and ions, as we shall discuss in the following.

At densities above $\approx 1.6 \times 10^{14} \, {\rm g \, cm^{-3}}$ there is a transition to a core fluid of neutrons, protons, electrons and possibly muons. The nature of the transition is uncertain. Some equations of state suggest that it is abrupt, while others predict a series of phase transitions in which nuclei are no longer spherical but arranged in rods and plates, the so called `pasta' phases in which matter behaves more like a liquid crystal than a standard crystal \cite{pasta}. The presence of such layers can have a strong observational impact on several phenomena (such as  glitches, oscillation modes of the star and gravitational wave emission)\cite{will} not only by lowering the shear modulus but also by smearing out the transition region. A perfectly solid crust presents, in fact, a sharp boundary for the interior fluid which reacts to changes in rotation by inducing a (dissipative) Ekman flow in a thin layer close to the boundary. The problem is complicated by the presence of magnetic fields and by neutron superfluidity but, generally, a smeared out boundary leads to weaker dissipation and longer crust-fluid coupling times.

In the outer core of the star protons are expected to form a type II superconductor\cite{BPPsuper} in which the magnetic flux is concentrated in flux-tubes. As we discuss below, the strong interaction between superfluid neutron vortices and flux tubes can lead to `pinning' also in the core. At even higher densities, above nuclear saturation density, the ground state of matter is essentially unknown. At asymptotically high densities the favoured state seems to be a condensate of deconfined quarks in the colour-flavour-locked (CFL) phase\cite{Alfcfl}. If such a phase exists (e.g. in some of the heavier stars) it may have interesting consequences; not only is it a superfluid but it also possesses a sizeable shear modulus\cite{CFLshear}, raising the possibility that `core quakes' contribute to pulsar glitches.

\subsection{Multifluid hydrodynamics}
\label{hydro}

In the previous section we purposely glossed over the details of superfluidity in the stellar interior and focussed on the constituents at different densities. Let us now consider in detail the consequences of superfluidity. We base our analysis on the variational formulation of Andersson \& Comer \cite{AC06} and consider the equations of motion for two dynamical degrees of freedom, two macroscopic `fluids': the inviscid neutron condensate and a viscous, charge-neutral fluid of protons and electrons. As already mentioned the neutron condensate rotates by forming an Abrikosov lattice of quantised vortices, which carry the circulation. Hence a hydrodynamical description only makes sense on length scales larger than, say, the mean intervortex separation, $d_v\approx 4\times 10^{-4} (P/1\mbox{ms})^{1/2}$ cm (see Haskell et al. 2012 \cite{HAC12} for an in depth discussion of the length-scales involved). Above this length scale we can average over the velocity field of the vortices contained in a fluid element.
If the stellar interior rotates uniformly, one can define an angular velocity for the condensate, which satisfies the relations:
\be
\kappa n_v=2[\Omega_\mathrm{n}+\varepsilon_\mathrm{n}(\Omega_\mathrm{p}-\Omega_\mathrm{n})]+\varpi\frac{\partial}{\partial \varpi}[\Omega_\mathrm{n}+\varepsilon_\mathrm{n}(\Omega_\mathrm{p}-\Omega_\mathrm{n})],\label{vortici}
\ee
where $\kappa=h/2m_n=1.99\times 10^{-3} \, {\rm cm^2\,s^{-1}}$is the quantum of circulation, with $m_n$ the neutron mass (in what follows we assume $m_n=m_p$, the proton mass), $n_v$ is the areal density of vortices, $\Omega_\mathrm{n}$ and $\Omega_\mathrm{p}$ are the angular velocities of neutrons and protons respectively, and $\varpi=r\sin\theta$ is the cylindrical radius.

If the individual species are conserved in beta equilibrium, we can write conservation laws for each species,
\be
\partial_t\rho_\mathrm{x}+\boldsymbol{\nabla}_i(\rho_\mathrm{x}v^i_\mathrm{x})=0,
\ee
with $\mathrm{x}=\mathrm{n},\mathrm{p}$. The Euler equations, on the other hand, take the form:
\be
(\partial_t+v_\mathrm{x}^j\boldsymbol{\nabla}_j)(v_i^\mathrm{x}+\varepsilon_\mathrm{x}w_i^{\mathrm{yx}})+\boldsymbol{\nabla}_i(\tilde{\mu}_\mathrm{x}+\Phi)+\varepsilon_\mathrm{x}w^j_\mathrm{yx}\boldsymbol{\nabla}_iv_j^{\mathrm{x}}=f_i^\mathrm{x}/\rho_\mathrm{x}+\boldsymbol{\nabla}^jD_{ij},\label{euler}
\ee
where $\varepsilon_\mathrm{x}$ is the relevant entrainment parameter (we recall that $\varepsilon_\mathrm{p}\rho_\mathrm{p}=\varepsilon_\mathrm{n}\rho_\mathrm{n}$), with $w_i^\mathrm{yx}=v_i^\mathrm{y}-v_i^\mathrm{x}$ ($\mathrm{y}\neq\mathrm{x}$), $D_{ij}$ represents the dissipative part of the stress-energy tensor (see Haskell et al. (2012) \cite{HAC12} for an in-depth discussion of dissipation in multifluid systems), $\tilde{\mu}_\mathrm{x}=\mu_\mathrm{x}/m_\mathrm{x}$ is the chemical potential and $\Phi$ is the gravitational potential, which obeys the Poisson equation:
\be
\nabla^2\Phi=4\pi G \sum_{x} \rho_\mathrm{x}.
\ee
The force $f_i^\mathrm{x}$ which appears on the right hand side of equation (\ref{euler}) is the so-called mutual friction force, which exchanges momentum between the components and is peculiar to a superfluid system. For straight vortices it takes the form:
\be
f_i^\mathrm{x}=\kappa n_v\rho_\mathrm{n}\mathcal{B}^{'}\epsilon_{ijk}\hat{\Omega}^j_\mathrm{ n}w^k_\mathrm{ xy}+\kappa n_v\rho_\mathrm{ n}\mathcal{B}\epsilon_{ijk}\hat{\Omega}^j_\mathrm{n}\epsilon^{klm}\hat{\Omega}_l^\mathrm{ n}w_m^\mathrm{ xy},\label{mutualF}
\ee
where $\mathcal{B}$ and $\mathcal{B}^{'}$ are defined as
\be
\mathcal{B}=\frac{\mathcal{R}}{1+\mathcal{R}^2},\;\;\;\mbox{and}\;\;\;\mathcal{B}^{'}=\frac{\mathcal{R}^2}{1+\mathcal{R}^2}.
\ee
$\Omega^i$ is the local angular velocity satisfying $v_i = \epsilon_{ijk} \Omega^j x^k$, and $\mathcal{R}$ is the dimensionless drag coefficient.

Mathematically $f_i^{\mathrm{x}}$ has the form of a body force acting on a fluid element, but its correct physical interpretation is subtly different. Consider again the most simple case relevant for neutron star interiors, that of a two fluid system of superfluid neutrons (denoted $\mathrm{n}$), coupled to a charge neutral fluid (denoted $\mathrm{p}$) of protons and electrons, which are locked by the Coulomb interaction on timescales much shorter than the dynamical timescales we are interested in \cite{prix04}. In this situation $f_i^\mathrm{x}$  arises from two distinct physical effects. First of all, in the presence of quantised vortices a new term appears in the hydrodynamical equations of motion for the neutron condensate. This term takes the same form as a (fictitious) Magnus force acting between the vortices and the condensate,
\be
f_i^M=\kappa n_v \epsilon_{ijk} \hat{\Omega}^j(v^k_\mathrm{n}-v^k_\mathrm{L}).\label{f1}
\ee
where $v_i^\mathrm{L}$ is the local velocity of a segment of vortex line (averaged over many vortices). Note that the Magnus term is only present if vortices are not `flowing' with the condensate. In the absence of additional forces that can lead to a relative flow between vortices and condensate (e.g. pinning or dissipation) the Magnus term thus vanishes.
Second there is a drag force which acts between vortices and the proton-electron fluid. It can be written phenomenologically as\cite{TrevMF}
\be
f_i^D=\kappa n_v \mathcal{R}(v^\mathrm{p}_i-v_i^\mathrm{L}),\label{f2}
\ee
where the exact microphysical process that give rise to dissipation will determine the value of the coefficient $\mathcal{R}$. In the case of straight vortices one can solve for $v_i^{\rm L}$ by assuming force balance  for massless vortices ($f_i^D=f_i^M$), yielding the expression in (\ref{mutualF}). 

It is important to note that the forces in (\ref{f1}) and (\ref{f2}) are 'average' forces acting on a small volume around a vortex. Given the scales in the problem it is possible to choose, for example, a cylinder such that it's radius is significantly larger than the vortex core but still small enough to allow us to consider the vortex as a 'line' in our hydrodynamical description. By integrating the momentum flux though the surface of such a cylinder one can obtain the force balance condition and derive the expression in (\ref{mutualF}) by enforcing linear momentum conservation. Strictly speaking, however, a vortex cannot be treated as a material object, as its centre is the location of a singularity of the flow, not a massive particle or body. The force is not localised on a vortex core, nor does it act equally on all fluid elements within the closed surface, as in a rigid body for example.  The acceleration of a vortex line is not well defined, so that only the average momentum transfer per unit time, and the average force balance condition $f_i^D=f_i^M$ are well defined.  An in depth discussion of how the Magnus force arises from the quantisation of vorticity is given, e.g. by Glampedakis et al. \cite{GAS}, Carter et al. \cite{CarterMagnus} and Sonin \cite{Sonin97}.

In the core of a neutron star dissipation is mainly from electrons scattering off vortex cores magnetised by entrainment\cite{AlparMF1} (see below) which leads to a drag parameter of the form\cite{TrevMF}:
\be
\mathcal{R}\approx 4\times 10^{-4} \left(\frac{\delta m^*_\mathrm{p}}{m_\mathrm{p}}\right)^2\left(\frac{m_\mathrm{p}}{m^*_\mathrm{p}}\right)^{1/2}\left(\frac{x_\mathrm{p}}{0.05}\right)^{7/6}\left(\frac{\rho}{10^{14}\mbox{g/cm$^3$}}\right)^{1/6},\label{mfcore}
\ee
where $m^*_\mathrm{p}=m_\mathrm{p}(1-\varepsilon_\mathrm{p})$ is the effective proton mass, one has $\delta m^*_\mathrm{p}=m_\mathrm{p}-m^*_{\mathrm{p}}$, and $x_\mathrm{p}=\rho_\mathrm{p}/\rho$ is the proton fraction. In the crust, where this mechanism does not act, as protons are not superconducting, the dissipation is mainly via phonon interactions \cite{PhononJones} or, for larger vortex velocities, via Kelvin waves excited along the vortices themselves \cite{KelvonJones, KelvonBaym}.
The values of the dissipation coefficients are, however, quite uncertain. They are likely to be low ($10^{-9} \lesssim \mathcal{R} \lesssim 10^{-7}$) for phonon scattering, while kelvon processes are strongly dissipative  ($10^{-3} \lesssim \mathcal{R} \lesssim 10^{-1}$) \cite{PhononJones, KelvonJones, KelvonBaym}.

The expression in (\ref{mutualF}) is appropriate for free, straight vortices and can easily be modified to account for the slight bending due to the finite rigidity of a vortex (see \cite{TrevTurb} for a description). However it is well known from the experimental study of He II \cite{Don} that a counterflow along the vortex axis destabilises the vortex array to form a turbulent tangle (the Glaberson-Donelly instability) \cite{DGunstable}. In this case the mutual friction can be described by the phenomenological Gorter-Mellink form\cite{GMMF}:
\be
f_i^{GM}=\frac{8\pi^2\rho_\mathrm{n}}{3\kappa}\left(\frac{\xi_1}{\xi_2}\right)^2\mathcal{B}^3 w_\mathrm{pn}^2w_i^\mathrm{pn},
\ee
where $\xi_1\approx 0.3$ and $\xi_2\approx 1$ are phenomenological parameters \cite{VinenMF}. Given that the Reynolds number in a neutron star\cite{melper07} is expected to satisfy $Re \gtrsim 10^{7}$, it is likely that instabilities occur, leading to superfluid turbulence and a vortex tangle \cite{Peraltaglitch}. Even at lower values of $Re$ it is well known from the study of laboratory superfluids that a turbulent vortex tangle is likely to develop \cite{Walmsley}.


\subsection{Vortex pinning}
\label{pin}

An important assumption in writing down (\ref{mutualF}) is that the vortices are free to move. However, this is not always so. Vortices can `pin' to nuclei in the crust or to magnetic flux tubes in the core of the star, thus preventing the neutron fluid from spinning down, as it cannot expel vorticity to keep up with the decelerating protons (to which we assume the magnetic field and observed radio pulses are anchored). The pinning force is thus an important ingredient of any superfluid glitch model, as it determines the amount of angular momentum that can be stored in the condensate, before the Magnus force eventually frees the vortices and gives rise to a glitch. 

\begin{figure}
\centerline{\includegraphics[width=9cm]{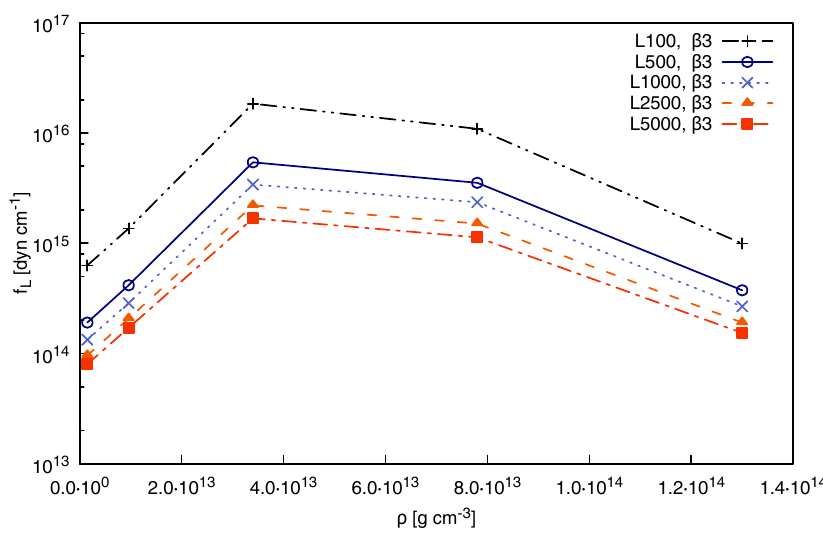}\;\;\;\includegraphics[width=9cm]{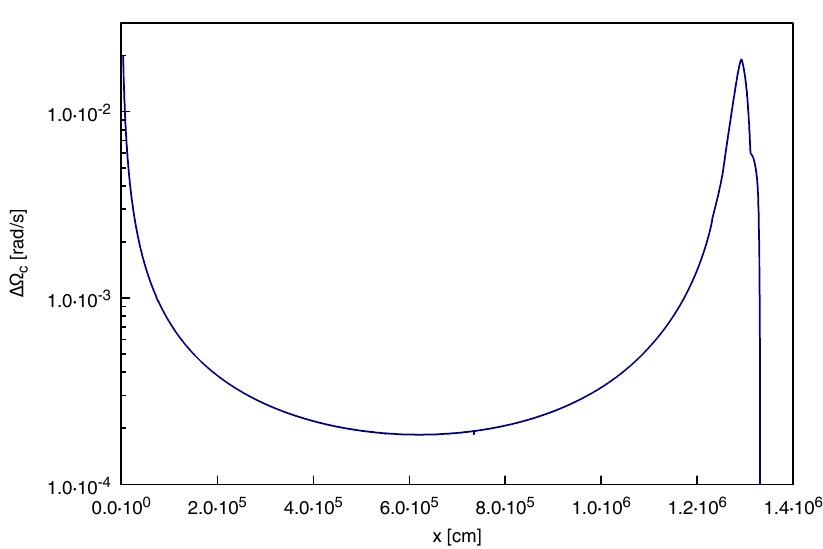}}
\vspace*{8pt}
\caption{LEFT: The pinning force at different densities obtained by Seveso et al \cite{Seveso14} for their more realistic $\beta=3$ model, which accounts for correlations in the superfluid pairing gaps. The results are shown for different values of the length scale $L$ over which a vortex can be considered rigid. RIGHT: Estimate, for a $1.4M_\odot$ NS modelled with the GM1 equation of state, of the critical difference in rotation rates between superfluid neutrons and the crust $\Delta\Omega=\Omega_\mathrm{n}-\Omega_\mathrm{p}$, at which vortices unpin at different cylindrical radii (from \cite{Seveso14}). \label{pinningF1}}
\end{figure}

Although the pinning force per pinning site is readily estimated, the pinning force per unit length acting on a vortex (which is the quantity that balances the Magnus force) is a more complex problem. The first quantity depends only on the difference in energy between the state where the nucleus is inside the vortex, and that where it is outside. The force per unit length on the other hand depends on several poorly constrained quantities such as the rigidity, orientation, and geometry (e.g. tangled versus straight) of the vortex among other things. Several early estimates \cite{AlparPin1, AlparPin2, BaymPin, RudPin} considered simple geometries and layers of different pinning strengths. It was also pointed out early on by Jones \cite{JonesNOPIN} that for an infinite vortex the energy differences between adjacent configurations are negligible once averaged over orientations, leading to a vanishing pinning force. Detailed calculations have been carried out by Link \cite{Link09}, who modelled the crust as a potential due to randomly placed ions, and by Seveso et al. \cite{Seveso14} who averaged over all orientations of the vortex with respect to a BCC crustal lattice.
Both calculations confirm that, for a finite vortex rigidity, the force is indeed reduced by the averaging procedure, but is still dynamically significant \cite{Seveso14, Link13}. In figure \ref{pinningF1} we show the results of Seveso et al. \cite{Seveso14} for the pinning force, also expressed in terms of the maximum lag between neutrons and protons that can be built up before the Magnus force breaks the vortices free. As we can see the strength of the pinning force spans two orders of magnitude across the inner crust, and has a maximum at densities around $\rho\approx (3-7)\times 10^{13}$ gm/cm$^3$, depending on the pairing model.


We can relate the critical lag in figure (\ref{pinningF1}) to the size of a glitch by a simple conservation of angular momentum argument.  Let us consider two rigidly rotating components: the crust and all components coupled to it on a short time-scale, with moment of inertia $I_\mathrm{p}$, rotating with angular velocity $\Omega_\mathrm{p}$; and a pinned condensate, with moment of inertia $I_\mathrm{n}$, rotating at $\Omega_\mathrm{n}$, with $\Omega_\mathrm{p}-\Omega_\mathrm{n}=\Delta$, where $\Delta$ is the critical lag. If all components recover fully after the glitch and rotate with common angular velocity $\Omega_a$, one has simply
\be
I_\mathrm{n}(\Omega_\mathrm{p}+\Delta)+I_\mathrm{p}\Omega_\mathrm{p}=(I_\mathrm{n}+I_\mathrm{p})\Omega_a~.\label{inertiaGG}
\ee
The size of a glitch,
$\Omega_a-\Omega_\mathrm{p}=-{I_\mathrm{n}}\Delta/{(I_\mathrm{n}+I_\mathrm{p})}$,
thus depends on the lag that builds up and on the fraction of the total moment of inertia that corresponds to the pinned superfluid. Note however that post-glitch corotation (i.e. erasure of $\Delta$) implies a correlation between glitch sizes and waiting times, which is not observed in most pulsars (see section 2). In this sense the above equation should be interpreted as an inequality. Essentially we are saying that in the case of strong pinning there is no vortex motion, so no mutual friction force acting between the components.

Let us examine the problem in more detail by writing the equations in (\ref{euler}) as \cite{Haskell12}:
\beq
\dot{\Omega}_\mathrm{n}&=&\kappa n_v \mathcal{B} \frac{(\Omega_\mathrm{p}-\Omega_\mathrm{n})}{(1-\varepsilon_\mathrm{n}-\varepsilon_\mathrm{p})}-\frac{\varepsilon_\mathrm{n}}{1-\varepsilon_\mathrm{n}}\dot{\Omega}_{sd}\label{system1}\\
\dot{\Omega}_\mathrm{p}&=&-\kappa n_v\frac{\rho_\mathrm{n}}{\rho_\mathrm{p}} \mathcal{B} \frac{(\Omega_\mathrm{p}-\Omega_\mathrm{n})}{(1-\varepsilon_\mathrm{n}-\varepsilon_\mathrm{p})}-\dot{\Omega}_{sd}\label{system2}\\
\kappa n_v&=&2[\Omega_\mathrm{n}+\varepsilon_\mathrm{n}(\Omega_\mathrm{p}-\Omega_\mathrm{n})]+\varpi\frac{\partial}{\partial \varpi}[\Omega_\mathrm{n}+\varepsilon_\mathrm{n}(\Omega_\mathrm{p}-\Omega_\mathrm{n})]\label{system3}
\eeq
where $\Omega_\mathrm{p}$ and $\Omega_\mathrm{n}$ may depend on position, if differential rotation builds up, and $\dot{\Omega}_{sd}$ is the observed spin-down rate, dictated by the external torque $T_{ext}$, which is generally assumed to be electromagnetic ($T_{ext} \propto \Omega_\mathrm{p}^3$). Let us first of all examine a simplified version of (\ref{system1})--(\ref{system2}) in which we neglect entrainment and the external spin-down torque (under the assumption that it acts on a much longer timescale than the mutual friction coupling) and let the fluids rotate rigidly. The solution is simply:
\be
\Omega_\mathrm{p}-\Omega_\mathrm{n} =\Delta_0 e^{-t/\tau}\;,\;\;\;\;\;\tau=\frac{I_\mathrm{p}}{I}\frac{1}{2\Omega_\mathrm{n}\mathcal{B}}
\ee
where $\Delta_0=(\Omega_\mathrm{p}-\Omega_\mathrm{n})$ is the initial lag in the pinned region, and the coupling time-scale $\tau$ depends on the ratio between the moment of inertia of the proton-electron fluid (and any component locked to it on a faster time-scale) and the total moment of inertia $I$. This is essentially  the two-component model proposed by Baym and collaborators soon after the observation of the first glitch\cite{2fluid}, in order to explain the slow relaxation observed after the glitch in terms of long superfluid coupling time-scales.

In neglecting the entrainment, however, we make a serious simplification. In the crust, $\varepsilon_n$ can be quite large \cite{Chamel1} (of order $\varepsilon_n\approx 10$). As can be seen from the right-hand side of equation (\ref{system1}), this leads to a spin-down torque acting on the pinned neutron condensate, even if we assume $\mathcal{B}\approx 0$ for strong pinning (i.e. zero drag force and mutual friction). This  strongly reduces the lag that can be built up between glitches and thus the amount of angular momentum that can be stored. To see this we can consider the accumulated activity of the Vela pulsar:
\be
\mathcal{A}=\frac{1}{t_{ob}}\sum_i \Delta\Omega_\mathrm{p}^i/\Omega_\mathrm{p},\label{Activity}
\ee
where the sum is performed over all the glitches and $t_{ob}$ is the observation timespan. With the aid of the simple angular momentum conservation argument presented above we can relate this quantity to the ratio between the moment of inertia of the pinned region and the total that is coupled during a glitch:
\be
\frac{I_\mathrm{n}}{I_\mathrm{p}}=-\frac{\Omega_\mathrm{p}}{\dot{\Omega}_\mathrm{p}}\mathcal{A} \;\;\;\;\mbox{{\bf without} entrainment}\label{inertia1}
\ee
which leads to $I_\mathrm{n}/I_\mathrm{p}=0.016$, a fraction which can be easily accommodated by the crust. However, recent work \cite{Chamel2, crustnot} has pointed out that in reality one has:
\be
\frac{I_\mathrm{s}}{I_\mathrm{c}}\approx-\frac{\Omega_\mathrm{p}}{\dot{\Omega}_\mathrm{p}}\mathcal{A}\frac{m_\mathrm{n}^*}{m_\mathrm{n}} \;\;\;\;\mbox{{\bf with} entrainment} \label{inertiaent}
\ee
which leads to a larger ratio $I_\mathrm{s}/I_\mathrm{c}\approx 4-6 \%$, using the results of Chamel \cite{Chamel1}. This ratio cannot be accommodated by the crust, unless the equation of state is soft and the star has mass $<1 M_\odot$. Alternatively it implies that the core is involved in the glitch, either by contributing to the pinning or by partially decoupling during and for some time after the event.

As a vortex approaches the threshold for unpinning, the probability rises that it hops to a nearby pinning site by thermal activation or quantum tunnelling \cite{LinkUnpin}, in what is known as vortex creep. Essentially one can split the interaction between a pinning centre and a vortex into a dissipative part, which we model as linear drag, as in the previous case, and a non dissipative part, $F_{pin}$. The equation for force balance takes the form \cite{HaskellLagrange}:
\be
\epsilon_{ijk} \hat{\Omega}^j(v^k_\mathrm{n}-v^k_\mathrm{L})+\mathcal{R}(v^\mathrm{p}_i-v_i^\mathrm{L})+F_{pin}=0,
\ee
where $F_{pin}$ is now the unspecified pinning force. The extra force allows extra freedom to redefine $\mathcal{B}$ and $\mathcal{B}^{'}$ independently\cite{Link12Crust}, although note that close to the pinning threshold the linear drag approximation breaks down\cite{Link09}.
We discuss creep in detail in section \ref{Creep}. However let us note that, in the linear regime which is mostly relevant for the short term relaxation of glitches, it can be incorporated in the formalism in (\ref{system1})--(\ref{system2}), by assuming that only a fraction $\gamma$ of vortices is free (i.e. are creeping) at a given time, with the others pinned. This leads to a drag force of the form $f^i_D=\kappa\gamma n_v \mathcal{R}(v^\mathrm{p}_i-v_i^\mathrm{L})$, which is equivalent to rescaling the coefficient $\mathcal{R}\longrightarrow\gamma\mathcal{R}$.
We see in the following sections that
\be
\gamma\approx e^{-\frac{E_p}{kT}\frac{(\Delta-\Delta_c)}{\Delta_c}},
\ee
where $\Delta=\Omega_\mathrm{p}-\Omega_\mathrm{n}$ is the lag, $\Delta_c$ is the critical lag for unpinning, $T$ is the temperature, $k$ Boltzman's constant and $E_p$ is the energy barrier that has to be overcome for unpinning (that in a first approximation is the pinning energy \cite{Alpar84a, lilaknock}).

Note that in the above we have assumed that the crust is completely rigid, either due to the rigidity of the lattice itself or due to the magnetic field coupling the charged components on short Alven timescales. This is a fairly good approximation on longer post-glitch relaxation timescales, but on shorter timescales additional dynamics is likely to be present, due to both elastic oscillations of the crust and Ekman pumping of the fluid, even in the presence of magnetic fields \cite{vE10,vE14}.

\subsection{Superconductivity}

Protons are expected to be superconducting in most of the core, with type II superconductivity preferred in the outer core, and a transition to type I superconductivity possible in the inner core\cite{GAS}.
For type I superconductivity the strength of the mutual friction is still in the range $\mathcal{R}\approx 10^{-4}$ \cite{Sedrakian1}. For type II superconductivity, the magnetic field is organised in flux tubes which interact with neutron vortices. The magnetohydrodynamic (MHD) equations for this situation have been derived by Glampedakis et al.\cite{GAS} and their main feature is that the usual Lorentz force is replace by a flux tube tension term. There is then in general an extra force between vortices and flux tubes that can lead to vortices pushing flux tubes out \cite{RUDcore, GAcore} and, more importantly, creates an energy barrier that can lead to pinning in the core. The threshold for unpinning in this case can approach\cite{Link03}
\be
\Delta\Omega_c\approx 1.5\times 10^{-2} \left(\frac{B}{10^{12}\mbox{G}}\right)^{1/2}\;\; {\rm rad\,s^{-1}}~.
\ee
Once cutting does occur, it is dissipative, with\cite{HaskellSat}
\be
\mathcal{R}\approx 2.5 \times 10^{-3} \left(\frac{B}{10^{12}\mbox{G}}\right)^{1/2}.
\ee
It could play a role in several areas, from glitch triggers \cite{RUDcore} to their relaxation \cite{LinkObserve, HaskellSuper} and even in limiting gravitational wave driven instabilities\cite{HaskellSat}.

\section{Glitch phenomenology}
\label{pheno}

Having discussed the interior physics of NS, let us turn our attention to the phenomenology of pulsar glitches. The spin frequency $\nu$ of a pulsar is measured by recording the time of arrival (TOA) at the telescope of each pulse and fitting a spin-down model obtained by Taylor expanding around a reference time $t_0$ :
\be
\nu(t)=\nu_0+\dot{\nu}_0(t-t_0)+\frac{1}{2}\ddot{\nu}_0(t-t_0)^2+...\label{preglitch}
\ee
After subtracting known systematic effects, such as the Earth's orbital motion and the proper motion of the pulsar, Equation (\ref{preglitch}) is generally a good fit to the data and higher order terms are not usually included. Slow deviations from Equation (\ref{preglitch}) are referred to as `timing noise', while sudden, impulsive, changes in the spin frequency and frequency derivative are glitches.

\subsection{Event statistics}

Glitches were first observed in the Vela and Crab pulsars in 1969 \cite{RM69, RD69,Crab1, Crab2} (soon after the discovery of pulsars themselves\cite{Pulsar}) and, while these remain some of the best studied systems, several hundred glitches have now been observed in over 100 pulsars \cite{Espinoza11}, including magnetars and millisecond pulsars.{\footnote{ An up to date list of glitches is maintained by Jodrell Bank Observatory and can be found at http://www.jb.man.ac.uk/pulsar/glitches.html} The growing volume of events allows one to disaggregate the data and highlight the following general trends \cite{Melatos08}}:
\begin{enumerate}
\item{{\bf Glitch sizes and waiting times.} Glitch sizes span several orders of magnitude, with the smallest glitches of the order of $\Delta\nu/\nu\approx 10^{-12}$  and the largest of the order of  $\Delta\nu/\nu\approx 10^{-5}$ (see e.g. Espinoza et al. \cite{Espinoza11}). Melatos et al. \cite{Melatos08} showed that the size distributions are consistent with power laws, with the index varying from pulsar to pulsar, and the waiting time distributions are consistent with exponentials, as one would expect from a scale-invariant or self-organized critical process, such as earthquakes or vortex avalanches \cite{Jensen}. There are, however, significant exceptions to this trend: in Vela, J0537-6910 and J1341-6220, glitches recur quasiperiodically \cite{Melatos08, George} and exhibit a narrower spread in size. This suggests the presence of an additional mechanism that sets some natural scale for the process and may set the maximum scale for glitches, although a rapidly forced critical system may also exhibit quasi-periodic avalanches, as we discuss in the following section. Very interestingly, a recent analysis of the lower end of the Crab's size spectrum, suggests the existence of a minimum size of $\Delta\nu/\nu\approx 10^{-9}$, which is significantly above the smallest resolvable event\cite{CrabSize}. The lower end of the distribution, however, is contaminated by a different population of timing noise events (possibly of magnetospheric origin) and further studies are required to identify a lower limit securely. At the upper end of the spectrum, across the entire population, Espinoza et al. \cite{Espinoza11} have identified a sub-class of `giant glitchers' that exhibit large glitches of similar sizes, $\Delta\nu\approx 10 \mu$ Hz.}

\item{{\bf Reservoir effect.} Do larger glitches occur after longer waiting times, i.e. does the star contain an angular momentum `reservoir' which is replenished before and is fully emptied during a glitch? Several models (e.g., star quakes) predict a size-waiting-time correlation, whereas others (e.g., self-organized critical systems) do not. Strikingly, such a correlation exists in the pulsar J0537-6910. However, the correlation does not exist in the rest of the population; in general, there appears to be no `reservoir' effect \cite{Melatos08}. In figure \ref{reservoir} we plot glitch size verses waiting time to both the previous and next glitch. Only J0537-6910 shows a correlation between the size of a glitch and the waiting time to the next, which can be interpreted as the time to reach a critical stress being proportional to the size of the previous glitch \cite{Mid06}.  Melatos et al. \cite{Melatos08} argued that the correlation observed in J0537-6910 is, in fact, a by-product of the observed quasi periodicity and the fact that large glitches, of approximately the same size, dominate the size distribution.}
\item{{\bf Age and glitching activity.} Approximately 10$\%$ of the pulsar population exhibits glitches, with seven pulsars having glitched more than 10 times . In general, middle-aged pulsars glitch the most; activity decays with age, and older pulsars have smaller glitches \cite{SL96}. There also appears to be a correlation between glitching activity and the spin down rate \cite{Espinoza11}. The Root Mean Square (RMS) variations in the residuals associated with timing noise also correlate positively with the spin-down rate \cite{SC10, LM14}. Electromagnetic spin down drives glitch activity in most models, so it is natural that older pulsars (which spin down slower) are less likely to glitch (let alone have large glitches) during the observations \cite{Melatos08}. 

\begin{figure}
\centerline{\includegraphics[width=9cm]{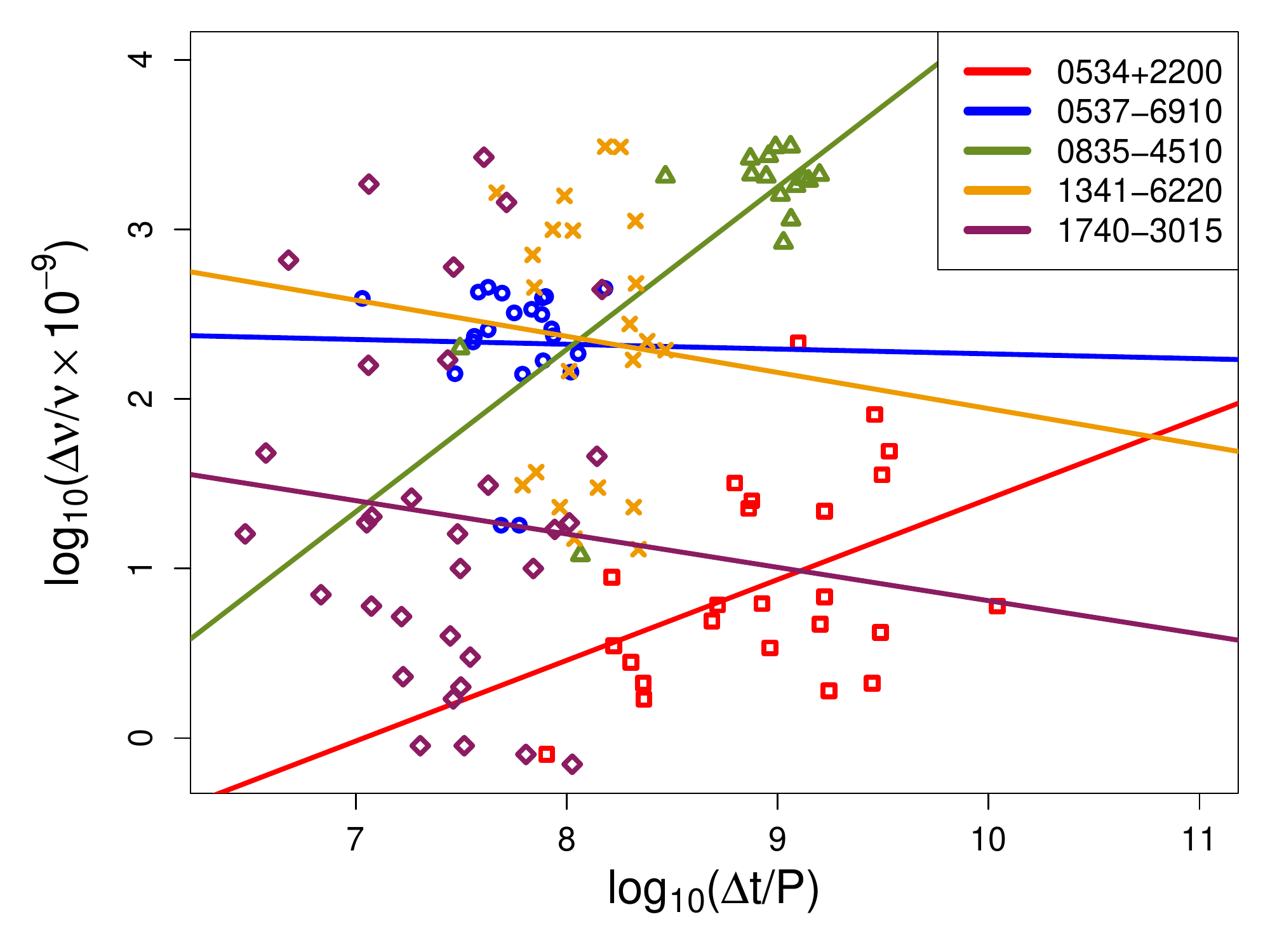}\;\;\;\includegraphics[width=9cm]{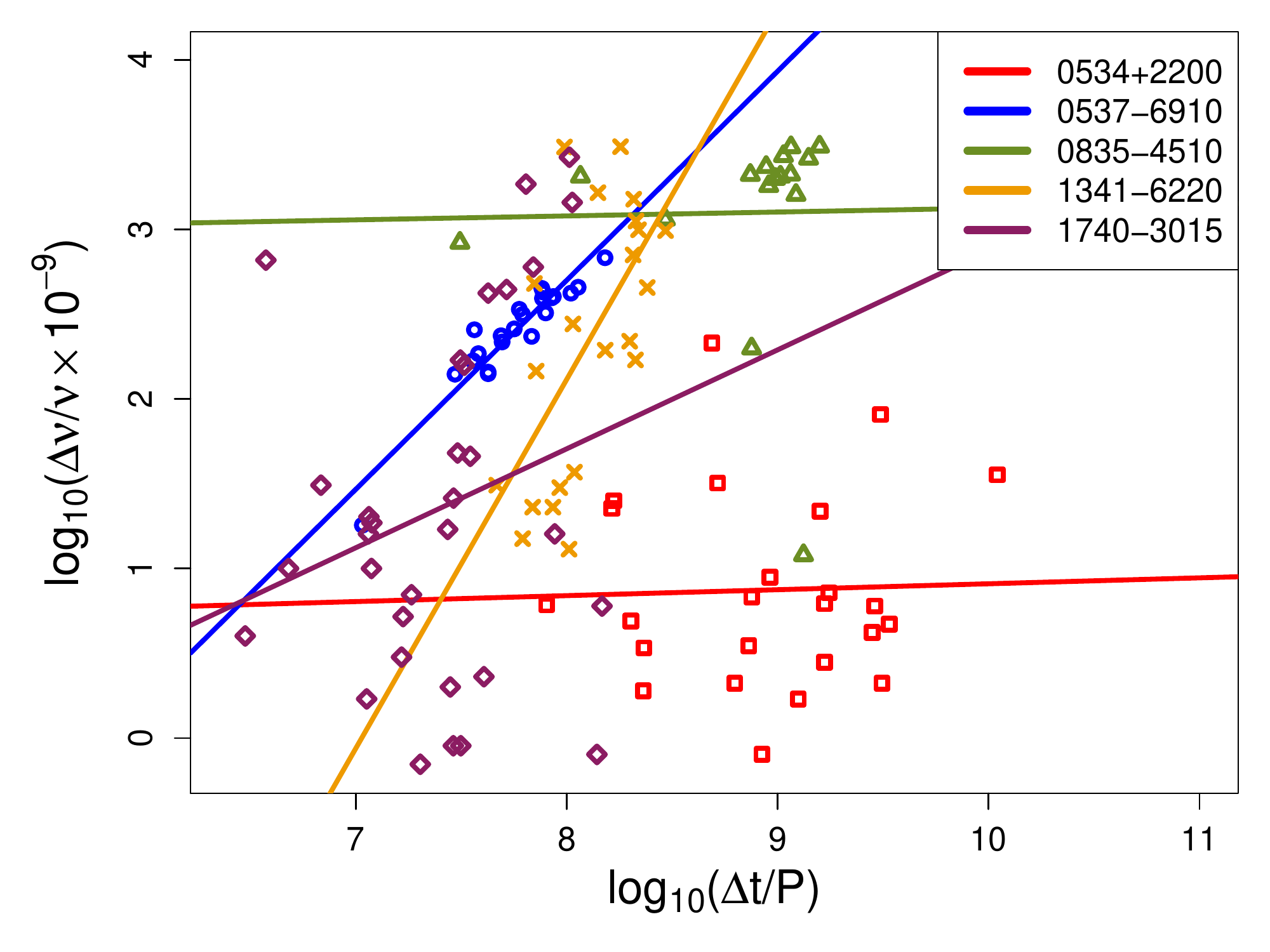}}
\vspace*{8pt}
\caption{LEFT: The correlation between glitch sizes and waiting times to the next glitch. RIGHT: The correlation between glitch sizes and waiting times from the previous glitch. The lines indicate the best linear fit and the correlation coefficients are shown in table \ref{clefs} \cite{George}.The only case in which there is a correlation is for the pulsar J0537-6910, in which there is a correlation between  the size of a glitch and the waiting time to the next, suggesting that the size of the glitch is random, but sets the time it takes to build up enough stress to reach the threshold again, as in the crust quake model.\label{reservoir}}
\end{figure}

\begin{table}
\caption{The correlation coefficients for the log-log fits in figure \ref{reservoir}, for correlations between glitch size and waiting time to the next glitch (case 1), and between size and waiting time front eh previous glitch (case 2). As can be seen on the pulsar J0537-6910 shows a significant correlation in case 1. }
{\begin{tabular}{l l l}
\hline
Pulsar &$\rho(1)$&$\rho(2)$ \\
\hline
J0537-6910&  $-$0.018 & 0.94\\
J0835-4510 (Vela)& 0.72 & 0.017\\
J0534+2200 (Crab) & 0.42 & 0.032\\
J1341-6220 & $-$0.062 & 0.63\\
J1740-3015 & $-$0.083 & 0.24\\
\hline
\end{tabular}}

\label{clefs}

\end{table}

    For pulsars that have glitched often another interesting quantity is the amount of the spin down that has been reversed by glitches, the activity parameter:
\be
\mathcal{A}=\frac{1}{t_{ob}}\sum_i \Delta\Omega_\mathrm{p}^i/\Omega_\mathrm{p},\label{Activity2}
\ee
As already discussed in the previous section this can be linked to the moment of inertia of the region initiating the glitch, from relations (\ref{inertia1}) and (\ref{inertiaent}). In table \ref{active} we show the activity parameter, and constraints on the moment of inertia of the glitching region, taken from Andersson et al. \cite{crustnot}. As we can see in general one needs at least a few percent of the total moment of inertia of the star to be transferred during glitches, possibly up to nearly $10\%$ if we account for the large neutron effective masses calculated by Chamel \cite{Chamel1}. }
\item{{\bf Radiative and pulse profile changes.} The general lack of correlation between glitches and radiative changes was taken early on as evidence for the internal origin of these phenomena. Nevertheless there are now some high field pulsars for which radiative changes, such as changes in the pulse profile, are observed in coincidence with glitches \cite{Lyne10, Liv11, Welte11, Keith13} as is the case for magnetars\cite{DK14}. It is thus clear that the magnetosphere plays a role in these systems, although how this could be, when the dynamical time-scale of the magnetosphere ($\sim$ spin period) is much shorter than the glitch waiting times, remains a mystery. In the case of magnetars, it is also thought that sudden changes in the magnetosphere are behind the observed phenomena of `anti-glitches', i.e., sudden, negative $\Delta\nu$. Correlated changes in the pulse profile and spin-down rate are also associated with timing noise events in several pulsars\cite{Lyne10}.}

A recent analysis of the Crab's spin-down rate and glitching activity has also claimed a link between an 11 year period of increased glitching activity, and changes in the breaking index in the same period, which would be linked to a change in the spin-down torque \cite{45crab}.
\end{enumerate}

\begin{table}
\caption{ We list, from  reference \cite{crustnot}, the characteristic spin down age $\tau_c$ and the activity parameter $\mathcal{A}$ for several pulsars, as well as the fraction of the total moment of inertia that is transferred during the glitch (the superfluid 'reservoir'), obtained from equation (\ref{inertia1}). Note that if one accounts for strong entrainment and uses equation (\ref{inertiaent}) to evaluate the ration of moments of inertia, the results are generally a factor 4-6 larger.}
{\begin{tabular}{l l l l}
\hline
Pulsar &$ \tau_c$ (kyr) & $\mathcal{A}$ ($10^{-9}$/d) & $I_\mathrm{n}/I$ (\%)\\
 & & &(no entrainment) \\
\hline
J0537-6910& 4.93 & 2.40 & 0.9\\
B0833-45 (Vela)& 11.3 & 1.91 & 1.6\\
J0631+1036& 43.6 & 0.47 & 1.5\\
B1338-62& 12.1 & 1.31 & 1.2\\
B1737-30& 20.6 & 0.79 & 1.2\\
B1757-24& 15.5 & 1.35 & 1.5\\
B1758-23& 58.4 & 0.24 & 1.0\\
B1800-21& 15.8 & 1.57 & 1.8\\
B1823-13& 21.5 & 0.78 & 1.2\\
B1930+22& 38.8 & 0.95 & 2.7\\
J2229+6114& 10.5 & 0.63 & 0.5\\
\hline
\end{tabular}}
\label{active}
\end{table}


\subsection{Post-glitch recovery}

The rise time of a glitch cannot be resolved at present. The best upper limit is $\approx 30$ s for the Vela pulsar if one neglects the two `slow glitches' observed in the Crab, which are thought to have been a different kind of event \cite{CrabSU, Wong01}. After the unresolved rise, there is a change in the spin-down rate, part of which is permanent, and part of which decays over days.  It is customary to fit these relaxation events with a function of the form:
\be
\nu(t)=\nu_{sd}(t)+\Delta\nu_p+\Delta\dot{\nu}_p t +\sum_{i=1}^N \Delta \nu_i e^{-t/\tau_i},\label{glitchsd}
\ee
where $\nu_{sd}(t)$ is the pre-glitch spin-down law in (\ref{preglitch}), $\Delta\nu_p$ is the `permanent' step in frequency, $\Delta\dot\nu$ is the `permanent' step in the spin-down rate, and $\Delta\nu_i$ are the amplitudes of the decaying parts of the frequency step, with decay time-scales $\tau_i$. The number of exponentials $N$ that is fitted varies from pulsar to pulsar and glitch to glitch and depends on how long after the glitch one has observations and on the sampling rate. For some of the better observed Vela glitches one needs to fit four exponentially decaying terms \cite{Dodson07}. The total size of the glitch, $\Delta\nu_G$, is
\be
\Delta\nu_G=\Delta\nu_p+\sum_{i=1}^N \Delta\nu_i.
\ee

The relaxation can vary significantly from pulsar to pulsar, and even from glitch to glitch in the same object. The latter property is especially surprising: glitches are small-amplitude events with $\Delta\nu_G/\nu\ll 1$, so one naively expects the recovery to be a linear process, whose time-scale is independent of its amplitude and depends on constitutive coefficients (e.g., viscosity and mutual friction) whose values do not change appreciably during inter-glitch intervals of years \cite{vE10}. In some objects the recovery is well approximated by simple steps in frequency and frequency derivative, but often this is due to infrequent observations that make it impossible to fit for short-term exponentials. Other glitches appear to be simple steps in frequency alone. Where $\Delta\dot{\nu}$ is actually measured, it is sometimes unclear whether one is measuring the enhanced spin-down rate during a long recovery, i.e., $\Delta\dot{\nu}\approx\Delta\nu_i/\tau_i$ or a genuine permanent torque change $\Delta\dot{\nu}_p$; it is a priority to clarify this issue observationally. Often different kinds of relaxation are observed in different glitches of the same object \cite{Yu, Espinoza11}.
Furthermore `microglitches' of small amplitude can have all kinds of signature, with negative and positive steps in $\nu$ and $\dot{\nu}$ \cite{Urama06, Urama10}. Recently a population of such events has been uncovered in the Crab pulsar \cite{CrabSize} and appears to be separate from the main glitch population and of likely magnetospheric origin (i.e., to be, in some sense, part of the timing noise).

The best resolved glitches are those of the Vela and Crab pulsar, for which we have the best coverage and largest body of data. The recoveries in these two NSs are quite different. For the more recent glitches of the Vela pulsar it is been possible to measure large increases in the spin-down rate, $\Delta\dot{\nu}/\dot{\nu}\lesssim 1$, which cannot be accounted for simply by assuming that part of the crust is decoupled but require changes in the internal torque \cite{JM05}. Moreover, if we define a healing parameter:
\be
Q=\sum \frac{\Delta\nu_i}{\Delta\nu_G},\ee
one finds $Q\gtrsim 0.8$ for the Crab \cite{Wong01}, and $Q\lesssim 0.2$ for Vela \cite{Crawford03}. In general $Q$ appears to correlate with $\dot{\nu}$ and young pulsars like the Crab exhibit near complete recovery and older pulsar like Vela only a very partial recovery \cite{LyneStats}. The measurements in Vela are complicated by the fact that a significant fraction of the glitch has not recovered before the next one occurs. It is interesting to note that in the Crab the  persistent offset in the spin-down rate ($\Delta \dot{\nu}_p$)leads to the star 'overshooting' and spinning slower than it would have had the glitch not occurred (an effect that may be masked in Vela by the occurrence of the next glitch). Link et al. \cite{Link92} argued that this is not possible in standard superfluid creep or crust quake models, unless one allows for a time-dependent torque that changes during the glitch (i.e., that is not purely a function of the lag between the crust and the superfluid) or if there are changes in the external torque, e.g., due to magnetic field rearrangement.
Van Eysden \& Melatos (2010) \cite{vE10} provided a different perspective, showing that the overshoot emerges naturally in a two-component superfluid model of glitch recovery featuring Ekman circulation and mutual friction, provided that the Ekman and mutual friction time-scales lie in a specified range. The van Eysden-Melatos model reproduces spin-down in laboratory helium experiments to better than $1\%$ \cite{vE12}.

The recoveries of magnetar glitches are also remarkable in many ways. Many exhibit large increases in the spin-down rate ($\Delta\dot{\nu}/\dot{\nu}\approx 0.5$) which persist over weeks to months \cite{Osso03, gavriil11}. Haskell \& Antonopoulou  \cite{HA14}  suggested that this is a  consequence of the long rotational period of these systems and that the physics governing the recoveries in magnetars is the same as in radio pulsars, although the glitch trigger may not be, with magnetospheric effects likely to play a role.
It has, in fact, been suggested that torque variations associated with magnetic field reconfiguration in the magnetosphere may be at the origin of the peculiar 'anti-glitch' that has recently been observed in the magnetar 1E 2259+586 by Archibald et al. \cite{antiG1, antiG2}. In this event there appears to be a sudden spin {\it down} (rather than up as in conventional glitches) of the star, followed by another timing event, which is harder to interpret, but is most likely a second anti-glitch \cite{antiG3}.

Upper limits on the temperature increase of the Vela pulsar have also been obtained from Chandra X-ray data, limiting the increase in temperature to $<0.2\%$ 35 days after the glitch \cite{heat1} and $<0.7\%$ 361 days after the glitch \cite{LarsLink}.

\subsection{Timing noise}

Timing noise is a general term to define stochastic deviations from equation (\ref{preglitch}). It manifests as noise in phase, frequency, or frequency derivative in different objects \cite{CD85, Ale06, Lyne06}. In general the effects can be due to various processes including precession \cite{Ale93, DR83}, Tkachenko modes \cite{Haskell11}, superfluid vortex creep \cite{alpartrigger}, superfluid turbulence \cite{LM14}, or changes in the magnetic field \cite{SL96, Urama06, Hobbs10}.
Timing noise is of interest for the current discussion as it has been suggested that glitches below the observational threshold may be the cause of timing noise \cite{alpartrigger, Cheng89} and of the measurement of anomalous breaking indices \cite{Ale95, Alpar06}. Janessen \& Stappers \cite{JS06} in fact showed that small glitches can account for part of the timing noise in PSR B1951+32, but not all of it, while Hobbs et al. \cite{Hobbs10} claimed that for young pulsars with $\tau_c<10^5$ yr timing noise is dominated by glitch recoveries. However recent measurements have highlighted correlations between pulse shape and timing noise \cite{Lyne10}, while surveys \cite{Lstat, Espinoza11, CrabSize} have uncovered a population of small, impulsive events that can be fitted as glitches as long as one allows for negative and positive $\Delta\nu$ and $\Delta\dot{\nu}$. This population appears to be distinct from the larger glitch population and is likely of magnetospheric origin.
If this is the case, glitch statistics are confusion-noise limited, as one reaches the level of sensitivity necessary to uncover timing noise  \cite{SC10, CrabSize}.

\section*{Triggers}

The rest of the review deals with the theories put forward to explain the observations summarized in Sec. 2. We divide the theories roughly into two groups: trigger mechanisms, which explain how the glitch originates in the first place and predict glitch sizes, waiting times, and their distributions; and relaxation mechanisms, which explain the shape and time-scales of the recovery. This distinction is convenient but somewhat artificial, as some theories address both issues, as discussed below.

\section{Crust quakes}

Soon after the first observations of glitches in the Vela and Crab pulsars, crust quakes were suggested as possible causes for these phenomena \cite{Rud69, Small70}. The fact that terrestrial earthquakes follow a power-law energy distribution makes star quakes an attractive explanation for the glitch size distribution observed. The crust forms a rigid lattice, which crystallized at a previous, faster, spin rate. As the star spins down, the crust maintains this shape, but it is no longer the equilibrium shape corresponding to the current angular velocity. Stress thus builds up. To quantify this process let us define the oblateness as \cite{BaymPines}:
\be
\epsilon = (I-I_{nr})/I_{nr},\label{oblate}
\ee
where $I$ is the actual moment of inertia of the star, and $I_{nr}$ is the moment of inertia of the star in the absence of rotation. The equilibrium value of $\epsilon$ is determined by the competition between gravity, which tends to make the star spherical, and rotation and elasticity, which tend to make it oblate.
The gravitational energy takes the form:
\be
E_g=E_{g}^{nr}+A\epsilon^2,
\ee
where $E^{nr}_g$ is the energy of the non-rotating configuration, and the coefficient $A$ is determined by the stellar structure; for an incompressible sphere one has $A=(3/25) (GM^2/R)$. The deviations are quadratic in $\epsilon$ as a sphere is the minimum energy configuration. The elastic energy can also be expanded around the unstrained state, that corresponds to a reference value $\epsilon_0$. In this case we have
\be
E_{el}=B(\epsilon-\epsilon_0)^2,
\ee
with $B=(57/50 \mu) (4\pi R^3/3)$, for a self gravitating sphere of constant shear modulus $\mu$, and, more generally, $B\ll A$. The rotational energy is
\be
E_{r}=L^2/2 I\;,\;\;\;\;\mbox{with}\;\;\;\;L=I\Omega,
\ee
where $L$ is the total angular momentum.
Upon minimizing the total energy $E_g + E_{el}+E_r$ while keeping $L$ constant, one finds
\be
\epsilon=\frac{I_{nr}\Omega^2}{4(A+B)}+\frac{B}{A+B}\epsilon_0.
\ee
The reference oblateness is
\be
\epsilon_0=\frac{I_{nr}\Omega_0^2}{4A},
\ee
with $\Omega_0$ the rotation rate corresponding to the relaxed state (i.e. the rotation rate at which the crust would be completely un-stressed). The mean stress in the crust is $\bar{t}=\mu(\epsilon_0-\epsilon)$
When a critical strain $\bar{\sigma}_c=\bar{t}_c/\mu$ is reached, the crust readjusts to a new reference shape (a crust quake) with a sudden negative change $\Delta\epsilon_0$. This shifts $\epsilon$ (the actual ellipticity of the star) by an amount:
\be
\Delta\epsilon=\frac{B}{A+B}\Delta\epsilon_0,
\ee
which leads to $\Delta\epsilon=\frac{\Delta I}{I_{nr}}=\frac{\Delta\Omega}{\Omega}$, where we use equation (\ref{oblate}) to relate the change in oblateness to a change (decrease) in moment of inertia and thus to the observed frequency jump $\Delta\Omega$.

Given the size of a glitch, the quake model predicts the size of the next event. The stress relieved during a quake is:
\be
\Delta \bar{t}=\mu(\Delta\epsilon_0-\Delta\epsilon)=\mu\frac{A}{B}\Delta\epsilon,
\ee
while the rate at which strain builds up is
\be
\dot{\bar{t}}=-\mu\dot{\epsilon}=\frac{\mu}{4A}\frac{I_{nr}\Omega^2}{\tau_{sd}},
\ee
where we define the spin-down age by $\tau_{sd}=-\Omega/\dot{\Omega}$. The waiting time to the next glitch is thus:
\be
\delta t_g\approx\frac{|\Delta\bar{t}|}{\dot{\bar{t}}}\approx\frac{4A^2}{B}\frac{\tau_{sd}}{I_{nr}\Omega^2}\frac{\Delta\Omega}{\Omega}.
\ee
Note that this assumes semi-complete release, which never happens in terrestrial earthquakes and Self Organised Critical Systems (SOCs) in general \cite{Jensen98, Melatos08}. Furthermore the change in moment of inertia produces a permanent change in the spin-down rate (assuming no change in the external torque)
\be
\frac{\Delta\dot{\Omega}}{\dot{\Omega}} \approx -\frac{\Delta I}{I}\approx -\frac{\Delta\Omega}{\Omega}.\label{odotcorrelate}
\ee

We can now compare the predictions to observations. The main difference from one star to another could, of course, be the factors $A$ and $B$, which depend on mass and radius. Nevertheless, for typical masses and radii ($M\approx 1.4 M_\odot$ and $R\approx 12$ km) the large glitches in the Vela pulsar would require almost all the stress accumulated in the crust from birth (assuming a period at birth of 1 ms) to be released during the glitch and furthermore would require such events to be rare, with waiting times of tens of millions of years. This is clearly not in accordance with what is observed; Vela has large, frequent glitches ($\Delta\Omega/\Omega\approx 10^{-6}$ approximately every 2-3 years). The situation is less severe for the Crab, but the core quake model would still require the Crab to have a low mass ($<1 M_\odot$)  \cite{BaymPines, PS72}. Equation (\ref{odotcorrelate}) also predicts smaller permanent steps in the spin-down rate than are observed in Vela, and a correlation between $\Delta\Omega$ and $\Delta\dot{\Omega}$ that is not generally observed in the pulsar population \cite{Melatos08}. There is also no observed correlation between the size of a glitch and the waiting time until the next, with the notable exception of PSR J0537-6910 (one of the quasi-periodic glitchers), where a  linear correlation has been claimed \cite{Mid06}.

The main ingredient missing from the quake picture is neutron superfluidity. Superfluidity plays an important role in building up stress in the crust, leading to quakes \cite{Rud76, ChamPin, Chau93}. This can be seen if we consider the average pinning force per unit volume acting on the lattice,
\be
F_i^P=n_v f_i(\varpi,z).
\ee
where $f_i$, the pinning force per unit length that opposes the Magnus force, is directed along the cylindrical radius ${\varpi}$ for a straight vortex. As $f_i$ depends on density, it varies along a vortex, leading to
\be
\epsilon^{ijk}\nabla_j F_k^P=n_v \frac{\partial}{\partial z} f^P_{\hat{\phi}} \neq 0.
\ee
Since $F_i^P$ shears the lattice it cannot be balanced by gravity, pressure or centrifugal forces, which are all expressed as gradients, but must be balanced by elasticity, building up a strain. The maximum lag that can be accumulated before the crust breaks can be estimated as\cite{Rud76}:
\be
\Delta\Omega\approx \bar{\sigma}_c \frac{\mu}{p}\frac{G M}{R^3\Omega}\;\;\mbox{s$^{-1}$},
\ee

with $\bar{\sigma}_c$ the critical breaking strain, $p$ the pressure and $\Omega$ the angular velocity of the star. If the crust is weak ($\bar{\sigma}_c\approx 10^{-5}$) one gets reasonable values of the lag, which could be reached in $\sim 3$ yr in Vela. However, the more realistic breaking strain, $\bar{\sigma}_c\approx 0.1$, obtained with molecular dynamics simulations \cite{Horowitz} implies that vortices unpin well before the critical strain is reached, ruling out crust quakes as a trigger mechanism for large glitches (but not for smaller glitches and in younger pulsars such as J05370$-$6910 or the Crab) \cite{alparS1, alparS2}. More detailed modelling of the breaking of the crust needs to be undertaken to confirm the nature of the fracturing process \cite{FrancoLink, Horowitz}. Carter \cite{Carter2000} showed that stress can build up even in the absence of vortex pinning, as long as the viscosity between the neutron superfluid  and the crust is low. In this case, if the neutrons rotate differentially, the crust must be strained to balance the Magnus force, and the mechanism proceeds as described above, generating star quakes once the breaking strain is exceeded. 

Ruderman \cite{Ruderman98} has suggested that stress may also build up as superconducting flux tubes are pushed out by vortices during the spin down of the pulsar. The footpoints of the flux tubes are attached to the crust and shear it, leading to a quake. During the quake, platelets may break off, carrying the frozen-in magnetic field and migrating towards the equator, where they are subducted. This reduces the magnetic dipole moment and hence the external torque, accounting  for a permanent change in $\dot{\nu}$. A detailed analysis, however, suggests that vortex motion would mainly wind up the field in the azimuthal direction \cite{GA11}. The model suffers from the same problems described above, namely that the crust would not break before vortices unpin (even if pinning is to flux tubes), and the crust cannot support cracks such as those needed for platelet migration \cite{JonesQuake, Horowitz}. Furthermore the model predicts that pulsars with $\nu<1$ Hz should not glitch (as essentially all the magnetic flux has already been pushed out, if they were born spinning fast), which is not the case.

Another possibility is that the crust is indeed too weak to be the seat of glitch activity, but that the star harbours a solid core which experiences `core quakes'  \cite{Rudermancore}. Recent calculations suggest that the core may, in fact, be composed of a quark condensate in the colour-flavour-locked (CFL) phase \cite{AlfordCFL}, with a shear modulus $\mu\approx 10^{33}-10^{34}$ erg/cm$^3$ \cite{Mannarelli}. The greater shear modulus and moment of inertia can lead to larger glitches and lower waiting times. In fact it has been noted by Alpar \cite{alparS2}  that one may expect larger glitches than are observed in this case. More work is needed to understand how a solid core fails upon reaching the critical stress, if weaker sections exist, the overall nature of the phase transition, and the predicted statistics.

\section{Vortex pinning and the 'snowplow' model}
\label{snowplow}

The standard view of pulsar glitches is that there is a superfluid component (usually assumed to be in the crust) whose vortices are pinned. Then, the superfluid cannot lose vorticity and spin down, lagging behind the normal fluid, which spins down electromagnetically along with the solid crust \cite{AI}. An important ingredient in this scenario is the maximum pinning force that the lattice in the crust (or flux tube array in the core) exerts on the vortices. Let us expand the discussion on the pinning force from section (\ref{pin}) . Recent estimates that account for finite rigidity and different orientations of the lattice allow a sizeable lag to build up, large enough to explain Vela glitches \cite{Seveso14}. Let us now analyse this scenario in more detail.

In his seminal work in 1970 Alpar \cite{AlparPin1} estimated the maximum pinning force acting on a nuclear cluster as:
\be
|F_p|\approx\left[n_G\frac{\Delta^2(n_G)}{E_F(n_G)}-n_0\frac{\Delta^2(n_0)}{E_F(n_0)}\right]\frac{V}{x}\label{pinalpar}
\ee
where $\Delta$ is the supefluid gap, $E_F$ the Fermi energy of the neutrons, $V$ is the volume of a nuclear cluster, $n_0$ is the neutron number density inside the clusters, while $n_G$ is the number density outside (in the free state), and $x$ is the scale of the interaction. One has $x=\mathrm{min}(R_{WS},\xi)$, where $R_{WS}$ is the radius of the Wigner-Seitz cell, $\xi=E_F/(k_F\Delta)$ is the vortex coherence length (the radius of the vortex core), and $k_F$ the Fermi momentum of the free neutrons. Essentially this is the difference between the energy cost of a vortex core of normal matter being placed at a density $n_0$ rather then $n_G$. If this is energetically favourable, with ${\Delta^2(n_G)}/{E_F(n_G)}>n_0{\Delta^2(n_0)}/{E_F(n_0)}$, vortices pin to clusters. To obtain the pinning force per unit length, to compare to the Magnus force, one simply divides by the lattice spacing $a=2 R_{WS}$. This estimate, and subsequent analysis such as that of Epstein and Baym \cite{EB88} who included the contribution of the kinetic energy of the superfluid, leads to rather large forces, of the order of $f_p=10^{18}$ dynes cm$^{-1}$, corresponding to a critical lag of $\Delta\Omega_c\approx 10$ rad s$^{-1}$. The critical lag is much larger than the lag that builds up between glitches ($\Delta\Omega\approx 0.01$ rad s$^{-1}$) for Vela, easily accommodating the angular momentum transfer needed to explain the glitch. It is also consistent with the lack of a reservoir effect in most pulsars, i.e. with the fact that  glitches do not appear to release all the angular momentum that has been stored since the previous event (see section \ref{pheno} for a detailed discussion). It requires a trigger mechanism that would release the stored angular momentum well before the lag corresponding to the maximum of the pinning force is exceeded. Several suggestions have been made, including quakes freeing vortices \cite{Pines80, Rud91} (although, as discussed in the previous section, it is unlikely that the crust yields before the vortices unpin) and pinning inhomogeneities leading to regions of vortex over-density and a locally perturbed Magnus force \cite{alpartrigger, Chau93}. Gross-Pitaevskii simulations do, in fact, show that over-densities of vortices are key drivers of vortex unpinning \cite{Lila12}.

The force obtained in equation (\ref{pinalpar}) is large mainly due to the simplified geometry, which assumes that the force per nuclear cluster acts over the whole vortex, essentially assuming that the vortex is aligned with one of the principal axis of the crustal crystal. Jones \cite{JonesNOPIN} pointed out that if one averages over all possible orientations of the crystal (effectively treating the crustal lattice as a random potential, which would also be the appropriate treatment if the crust is an amorphous solid), the energy difference between the pinned and unpinned states depends on the fractional difference in the number of nuclei threaded by the vortex in the two states, leading to a vanishing pinning force in the limit of infinitely long vortices.

The length scale over which a vortex can bend to accommodate pinning sites has been estimated by Link and Cutler \cite{LinkCutler} as $l\approx (10^1-10^2) R_{WS}$. Seveso et al. \cite{Seveso14} obtain a slightly larger value of  $l\approx (10^2-10^3) R_{WS}$ and a different functional dependance on the tension and pinning energy, in agreement with the calculations of Hirasawa and Shibazaki \cite{HST}. Given the length scale over which a vortex can be taken to be rigid, one can calculate the pinning force per unit length. This has been done by Link \cite{Link09, Link13}, who simulated vortex motion in a random potential and obtained a pinning force per unit length $f_p$ of the form
\be
f_p\approx \frac{2 E_p}{R_{WS}\xi}\left(\frac{4 E_p}{3 R_{WS} T_v}\right),
\ee
where $E_p$ is the pinning energy and $T_v$ is the self energy (tension) of a vortex. For typical numbers this gives forces of the order $f_p\approx 10^{16}$ dynes cm$^{-1}$.

Recently a similar value was obtained by Seveso et al. \cite{Seveso14}, who calculated the pinning force per unit length of a vortex, using the superfluid gaps of Donati \& Pizzochero \cite{DP}. In this case the the force is obtained by integrating over all vortex orientations  the quantity 
\be
|f_p|(\theta,\phi)=|F_p|\frac{\Delta n(\theta,\phi)}{l},
\ee
where $\Delta n(\theta,\phi)$ is the difference in number of intercepted nuclear clusters, as a function of the angles $(\theta,\phi)$ which represent the orientation of the vortex with respect to a reference position. Averaging over all possible orientation reduces the force considerably, leading to forces in the range  $f_p\sim 10^{15}$ dynes cm$^{-1}$, and critical lags of $\Delta\Omega\sim 0.01$ rad s$^{-1}$. The exact value depends on the region of the crust considered and weakly on the equation of state and stellar model. Figure (\ref{pinningF1}) shows an example of a realistic density dependence of the lag, which clearly has a maximum in the strongest pinning regions. These more realistic values are comparable to the lag that Vela can build up between glitches, and the appearance of a maximum in the equatorial region, allows for a new picture of glitches for pulsars which exhibit a reservoir effect, the so-called `snowplow' model \cite{pierre}.

In the snowplow model vortices unpin in the stellar interior, as the threshold for unpinning is exceeded, and are assumed to eventually repin in the stronger pinning regions at the equator, gradually moving outward as a larger lag builds up. The model assumes that a glitch occurs once the critical lag corresponding to the maximum of the pinning force is reached. At this point the vortices are distributed in a thin sheet close to the region in which the pinning force peaks, containing $N_v=(2\pi/\kappa) r_{max}^2\Delta\Omega_{max}$ vortices, where $r_{max}$ is the cylindrical radius at which the maximum $\Delta\Omega_{max}$ of the critical lag is located. Once the maximum is exceeded vortices are free to move out an annihilate at the edge of the neutron drip point $R_{nd}$, leading to an exchange of angular momentum:
\be
\Delta L_{gl}=2\kappa N_v \int_{r_{max}}^{R_{nd}}  \mathrm{d}x  x\int_0^{l(x)/2} \mathrm{d}z \rho_\mathrm{n} (\sqrt{x^2+z^2}).,
\ee
where $\rho_\mathrm{n}(r)$ is the radial density profile of the superfluid neutrons. To obtain the physical observables of the glitch one must make assumptions about the superfluid fraction of the moment of inertia $Q_n=I_\mathrm{n}/I$  and the fraction of total moment of inertia that remains coupled  to the 'normal' component of the crust during a glitch, denoted $Y_{gl}$. The step in frequency is then given by\cite{pierre}
\be
\Delta\nu=\frac{\Delta L_{gl}}{2\pi I} \frac{1}{1-Q_n(1-Y_{gl})},\label{jump}
\ee
and the step in frequency derivative, immediately after the glitch, by
\be
\frac{\Delta\dot{\nu}}{\dot{\nu}}=\frac{Q_n(1-Y_{gl})}{1-Q_n(1-Y_{gl})}.\label{jump2}
\ee
The waiting time between glitches is simply given by the time to build up the maximum lag again after a glitch:
\be
\tau_g=\frac{\Delta\Omega_{max}}{\dot{\Omega}}\label{wait}
\ee
By fitting the above expression to the average waiting time of Vela glitches (which we remind the reader is a quasi-periodic glitcher) one can then fix the amplitude of the pinning force (up to uncertainties in the vortex tension) and then use (\ref{jump2}) to estimate $Y_{gl}$, having calculated $Q_n$ from a set of representative equations of state \cite{SPH12}. By comparing the predictions to Vela glitches one can then estimate that less than $10\%$ of the moment of inertia must be decoupled during the glitch. By comparing equation (\ref{jump2}) to the short term increase in spin-down rate after the 2000 and 2004 glitches one can constrain the gap model used for the calculations \cite{SPH12}. A comparison of equation (\ref{wait}) to the average waiting times for giant glitches in the sample of Espinoza et al. \cite{Espinoza11} reveals good agreement also for these systems \cite{Haskell12}.

Despite these encouraging results there are some issues with this model. The first is that the equations in (\ref{jump}-\ref{wait}) do not account for entrainment in the crust \cite{Chamel1, Chamel2}, which strongly reduces the lag, and hence the amount of angular momentum that can be exchanged. This effect can be parametrised in the model by allowing for larger values of $Y_{gl}$. However these are not easy to reconcile with the constraint in ($\ref{jump2}$) and the activity of Vela \cite{NewtonGlitch}. Haskell et al. \cite{Haskell12} simulated the hydrodynamical response of the star to a glitch triggered by the snowplow mechanism and showed that, for some hours after the glitch, the main contribution to the increased spin down rate comes from the mutual friction in the outer core, not from the external torque. The constraint in (\ref{jump2}) is thus not relevant for the observed short-term increase in $\dot{\nu}$ of the Vela pulsar. Haskell et al \cite{HaskellSuper} also show that to fit the post-glitch relaxation in Vela one cannot allow for part of the vorticity in the core to remain pinned to flux tubes after the glitch, as this would lead to the crust relaxing slower than observed.

A more important issue is that the snowplow model predicts glitches to be periodic, contrary to what is observed in most pulsars. So while the model can explain the glitching behaviour of Vela and the other quasi-periodic glitchers, it fails to account for other systems in which the size distribution is well described by a power law and the waiting time distribution by an exponential \cite{Melatos08}. In general it is thus expected that, while the snowplow model can set the maximum size of glitches in a system, other mechanisms must be active to account for the smaller, more frequent glitches. Quakes may be responsible, as already discussed, but an intriguing alternative is that vortices move collectively in avalanches, as we now describe.

\section{Scale invariance and Vortex Avalanches}

The apparent power law statistics of pulsar glitch sizes, spanning up to four decades in individual pulsars, point to a scale invariant process. One possibility that is especially pertinent to a slowly driven vortex array is Self Organised Criticality (SOC). Self organised criticality  was first invoked by Bak \cite{Bak1, Bak2} to explain how simple physical rules can lead to complex behaviour in sub-critical systems (i.e. systems which are marginally stable and close to an instability threshold). Examples of SOC can be found in several fields of physics, ranging from earthquake and forest fire modelling to superconductors \cite{Jensen}.
Warszawski and Melatos \cite{Lila08, Lila11, lilaknock} have shown that pulsar glitches may by a manifestation of SOC, which would explain many aspects of the phenomenon and describe how collective motion of vortices can trigger glitches of varying size. In particular two different mechanisms have been considered as the source of scale-invariant dynamics in pulsars: vortex avalanches \cite{Lila11, lilaknock} and coherent noise processes \cite{Lila09}. Let us analyse both these mechanisms in more detail.

\subsection{{SOC and Nearest-Neighbour Avalanches}}

A system in an SOC state is generally composed of many discrete elements, whose motion is dominated by nearest neighbour interactions rather than by global forces. Each element moves when the {\it local} force exceeds a threshold, thus allowing for stress to accumulate in certain regions and to relax rapidly in others. The system evolves through a series of metastable states, driven by an external driver which operates on a much slower time-scale than that associated with the local interactions. Over long time-scales, the system self-adjusts to fluctuate around a `self-tuned' configuration which is at the threshold of relaxation, e.g., a sandpile tends to the slope at which sand grains start to topple \cite{Jensen98}. In the latter sense, the system is `critical.'

This definition resembles quite closely our understanding of superfluid vortex dynamics in a pulsar crust. Electromagnetic spin down drives the system on long timescales. Vortices unpin when the Magnus force exceeds a local threshold, and interact with each other locally on a fast time-scale. The critical state here is exactly analogous to the Bean state in a type II superconductor \cite{Field95}: a vortex array which is just irregular enough such that each vortex is on the threshold of unpinning. (Note that this is not a regular Abrikosov lattice, where the Magnus force is zero everywhere). Alpar suggested early on \cite{alpartrigger, self, capacitor} that there may be capacitive and depleted regions of vorticity in the crust, in which steep gradients in the vortex density produce an increased Magnus force. Random unpinning may then lead to a vortex avalanche via knock-on effects. In fact, this is exactly what is observed in SOC systems. In general transitions from one metastable state to the next occur via avalanches that have no preferred scale and lead to distributions of sizes and lifetimes (durations of the avalanches) that are power laws.

To investigate this scenario quantitatively Warszawski and Melatos \cite{Lila08} constructed a two-dimensional cellular automaton, containing $N^2_c$ cells covering an equatorial section of a star of radius $R$. To make the problem tractable, vortices are grouped in bundles, which are assumed to be the discrete elements of the simulation. In general there are $N_b=N_v/(B_bN_{c}^2)$ elements, where $B_b$ is the average number of bundles per cell, and $N_v$ is the total number of vortices in the star, obtained from equation (\ref{vortici}). Note that only $\epsilon N_v$ vortices are assumed to be pinned; the rest keep up with the externally induced spin down of the crust. This feature of the model is unrealistic, at least in the inner crust, where pinning sites are separated by $\lesssim 10^2$ fm. One then defines a maximum pinning force, which gives a threshold Magnus force for unpinning $F_{th}$, assumed to be uniform. For each cell one then compares this threshold to the Magnus force acting on the bundles, which is calculated as
\be
F_M^i=\rho_\mathrm{n}\epsilon^{ijk} \kappa_j (v^v_k-v^\mathrm{n}_k),
\ee
where $v^v_i$ is the vortex velocity (equal to the crust velocity, as the vortices are pinned), and the velocity of the superfluid neutrons is calculated from
\be
v_\mathrm{n}^i=v^i_\mathrm{pinned}+v^i_\mathrm{unpinned}+v^i_\mathrm{NN},\label{Magnus}
\ee
where the first two terms are due to the pinned and unpinned vortices contained within a circle of radius $r_{cell}$, the position of the cell, and calculated from the Feynman relation in equation (\ref{vortici}). The last contribution is that of the eight nearest neighbouring cells. The system is evolved in time according to the following rules:
\begin{enumerate}
\item{The grid is initialised by placing vortex bundles at random.}
\item{The Magnus force is calculated according to the prescription in (\ref{Magnus}) and compared to the threshold $F_{th}$. Cells in which the threshold is exceeded are marked as supercritical.}
\item{Half the bundles in supercritical cells are moved to adjacent cells, following the direction indicated by the Magnus force imbalance $F_M^i-F_{th}^i$.}
\item{The number of unpinned bundles is recorded.}
\item{Steps (1)--(4) are repeated until no more supercritical cells are present.}
\item{One of two things is done: in the spin down case the spin frequency of the crust (and thus $|v_v|$) is decreased by $\dot{\nu}\Delta t$, with $\Delta t$ the length of the time step, and the number of background unpinned vortices is proportionally reduced. Otherwise the authors mimic thermal creep by redistributing bundles in random cells to their neighbours.}
\end{enumerate}

In general, the system requires fine-tuning of $\epsilon$ and $F_{th}$ to obtain scale-invariant avalanches. In this sense, the automaton does not strictly display SOC dynamics; it does not fine-tune itself \cite{Lila08}. Once stationarity is achieved, the waiting times $\tau_g$ are exponentially distributed and the distribution of sizes is indeed a power law, with exponent $b$ satisfying $2.2\lesssim t b \lesssim 5.5$. Strong thermal creep leads to deviations from a power law and an excess of small events. In general, no correlation is observed between waiting times and the size of glitches, in line with pulsar data (see Sec. 2). The rise times of the glitches (i.e. the duration of the avalanches) are also distributed as a power law.

In general, the behaviour of the system reproduces the observational data well and gives an explanation for the collective behaviour of vortices. Furthermore the assumption that nearest neighbour interactions dominate appears to be confirmed, at least on a small scale, by quantum dynamical simulations of vortex motion, as we discuss in Sec. 5.3. Naturally, the automaton above is simplified in many ways. For example the pinning threshold is taken to be uniform, so only an annular shell (`critical circle') in the star is near the unpinning threshold. Given the importance of this active region, a more detailed analysis should include a realistic density dependence of the pinning force, such as that used in the snowplow models, (see Sec. 4 for a description of how the active region is defined in these models), and a fuller study of how the critical circle changes radius as the pulsar ages. Further work is also needed to quantify the effect of the bundling parameter $B_b$. As $B_b$ increases, the size distribution develops a bump which is suggestive of  a system-size effect.

\subsection{Coherent Noise and Variable-Pinning Avalanches }

The main difference between avalanches triggered by coherent-noise and nearest-neighbour processes is that the former occur in response to changes in the global, external driver and occur even in the absence of large-scale spatial inhomogeneities. The physics is as follows. If the pinning thresholds span a range, vortices unpin readily from low-threshold sites and `stick' in high-threshold sites over time, developing temporal correlations which lead to scale invariant behaviour, even when the system is spatially uniform. In the pulsar case the driver is the Magnus force that is built up by spin down. The pinning sites are assumed to be abundant and homogeneously distributed on macroscopic scales (i.e. we do not assume the existence of large scale capacitive regions with stronger pinning, as postulated in Sec. 5.1 and  Alpar et al \cite{alpartrigger, capacitor}). Consider, for example, an ensemble of pinning sites (that are assumed to be defects in the nuclear lattice) each denoted by pinning strength $F_p^i$, where $i$ labels the site. The pinning strengths are drawn from a top-hat distribution $\phi(F_p)$ with mean $F_0$ and halfwidth $\Delta$, such that
\be
\phi(F_p)=(2\Delta)^{-1} \Theta (F_p-F_0+\Delta)\Theta(-F_p+F_0+\Delta)
\ee
where $\Theta$ is the Heaviside step function. Each one of $N$ pinned vortices occupies initially the site of a defect, and the system is evolved in time with the aid of an automaton following four simple rules \cite{Lila09}:
\begin{enumerate}
\item{A value of the global Magnus force $F_M$ is drawn from a distribution $\psi(F_M)$ which is assumed to have the same form as the observed glitch waiting time distribution, for a pulsar with mean glitch rate $\lambda$:
\be
\psi(F_M)=\sigma^{-1} \exp{(-F_M/\sigma)}
\ee
with
\be
\sigma=2\pi\dot{\nu}R\rho_\mathrm{n}\kappa/\lambda \label{sigma}
\ee
$F_M$ may be regarded as a measure of the volume-averaged Magnus force that builds up before a glitch. In a more realistic model, equation (\ref{sigma}) would involve not just $\lambda$ but also $(1-\delta\nu/\delta\nu_{cr})$ where $\delta\nu$ and $\delta\nu_{cr}$ are fluctuating and critical crust-superfluid lags respectively. }
\item{A small fraction $f\ll 1$ of vortices are allowed to unpin at random due, for example, to thermal fluctuations or creep.}

\item{The stress $F_M$ is applied to all the remaining $(1-f)N$ pinned vortices, which are allowed to unpin if $F_p^i<F_M$.}
\item{Each unpinned vortex repines at a nearby defect $j$ and is assigned a new threshold $F_p^j$.}
\end{enumerate}
Step (2) is necessary to avoid the system settling in an equilibrium with all vortices accumulating in ever stronger pinning sites [if $\phi(F_p)>0$ as $F_p\rightarrow\infty$], such that unpinning becomes rarer as time progresses. A more elaborate model that allows for nearest neighbour interactions would not have this requirement and could produce scale invariance even in the absence of thermal creep.

Warszawski and Melatos \cite{Lila09} found that this model reproduces aspects of the observed glitch population. For example, it gives rise to a power law distribution of sizes with exponent ranging from zero to two. Fits to the seven most active Poisson-like glitchers suggest $0.2\lesssim F_0/\sigma\lesssim 3$ and $0.1\lesssim\Delta/F_0\lesssim 0.9$. The main predictions of this model that are different from the vortex avalanche model are:
\begin{itemize}
\item{ The presence of aftershocks, due to the fact that the system tends to populate sites with larger $F_p^i$ as time passes. After a large glitch many sites are repopulated at random, leading to a lower average $\langle F_p\rangle$ in occupied sites, and thus to a higher probability of another glitch.}
\item{ Glitch rise times are also distributed as a power law with exponent $\approx 1$. This prediction is of course very difficult to test, as the glitch rise time is too short to be resolved by current radio observations.}
\item{ There is a correlation between sizes and waiting times, which is absent in the avalanche model. Such a correlation would be erased to some extent (to be determined by future studies) if the coherent-noise process is combined with nearest-neighbour interactions (Sec. 5.1) in a more realistic model as foreshadowed above.}
\item{ Spikes may be present at the upper and lower ends of the size distribution for $\Delta\ll\sigma$, $F_0$, i.e., for $\phi(F_p)$ narrow. The absence of the spikes in data suggests either $\Delta\approx F_0$ or that radio timing experiments do not resolve the smallest glitches. Spikes also occur, in principle, if $\psi(F_M)$ has a quasiperiodic term proportional to $\delta(F_M-F_0)$, but they do not contain much integrated probability; it is hard to extract much significance from their absence in the data.}
\end{itemize}

\subsection{Gross-Pitaevskii simulations}

A cellular automaton finesses the (huge) challenge of simulating the collective behaviour of the $\approx 10^{18}$ vortices which thread a NS interior. However, the choice of the `rules' that govern the automaton can be informed by what smaller-scale simulations teach us about the collective behaviour of vortices. 

To study this problem Warszawski and Melatos \cite{Lila11} evolve a two-dimensional, decelerating, pinned superfluid by solving the two-dimensional dissipative Gross-Pitaevskii equation (GPE)
\be
(i-\gamma)\hbar\frac{\partial\psi}{\partial t} = -\frac{\hbar^2}{2m}\nabla^2\psi - (\mu-V - g|\psi|^2)\psi-\Omega\hat{L}_z\psi,
\ee
where $\Omega$ is the angular velocity of the frame (i.e. of the crust), $\psi(x^i,t)$ is the superfluid order parameter, $V(x^i)$ the external potential (which represents both the container and the pinning sites), $\mu$ the chemical potential; $g$ parametrises the strength of the self interaction, and the angular momentum operator is given by $\hat{L}_z=-i\hbar\hat{z}\partial/\partial\phi$. The term $-\gamma\partial\psi/\partial t$ is a phenomenological dissipative term.

The GPE  accurately describes systems with weak interactions, such as Bose Einstein Condensates (BECs), and has been successfully used for a variety of analytical and numerical studies. It is debatable whether the nuclear many-body forces in a neutron star are `weak' in this sense; more work is needed on this question. To represent the typical situation in a NS (`crust') the potential is taken to be a  trapping potential which represents the container and several `spikes' to represent the pinning sites, of the form:
\be
V=V_\mathrm{trap}+\sum_i V_i[1+\tanh (\Delta|r-R_i|)],
\ee
where $V_\mathrm{trap}$ is the background potential, $R_i$ is the position of the $i$th pinning site and $V_i$  and $\Delta$ parametric their strength and width. The response of the container to vortex movement is calculated according to:
\be
I_c\frac{d\Omega}{dt}=-\frac{d\langle\hat{L}_z\rangle}{dt}+N_{EM},
\ee
where $I_c$ is the moment of inertia of the crust and $N_{EM}$ is the external electromagnetic spin down torque.

\begin{SCfigure}
\centering
{\includegraphics[width=9cm]{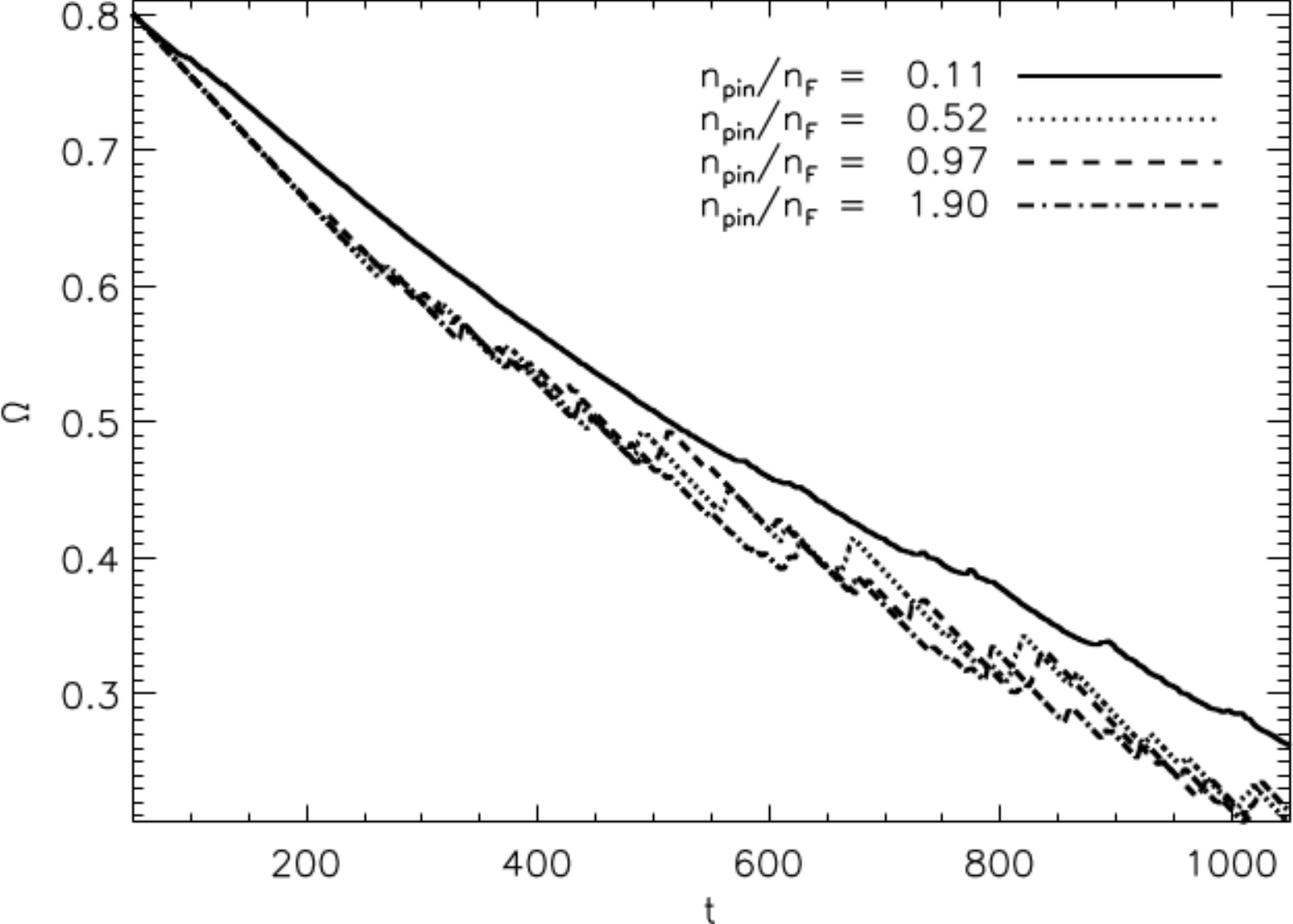}}\;\;\;\;\;\;\;\;
\caption{Example of results for the spin down of the crust in different numerical experiments, with varying ratio of occupied to free pinning sites, $n_{\mathrm{pin}}/n_\mathrm{F}$, as described in Warszawski \& Melatos (2011) \cite{Lila11}. One can clearly see that the spin-down is more spasmodic for a higher number of pinned vortices.\vspace{15pt}} \label{lila1}
\end{SCfigure}
\begin{figure}
\centerline{\includegraphics[width=8.9cm]{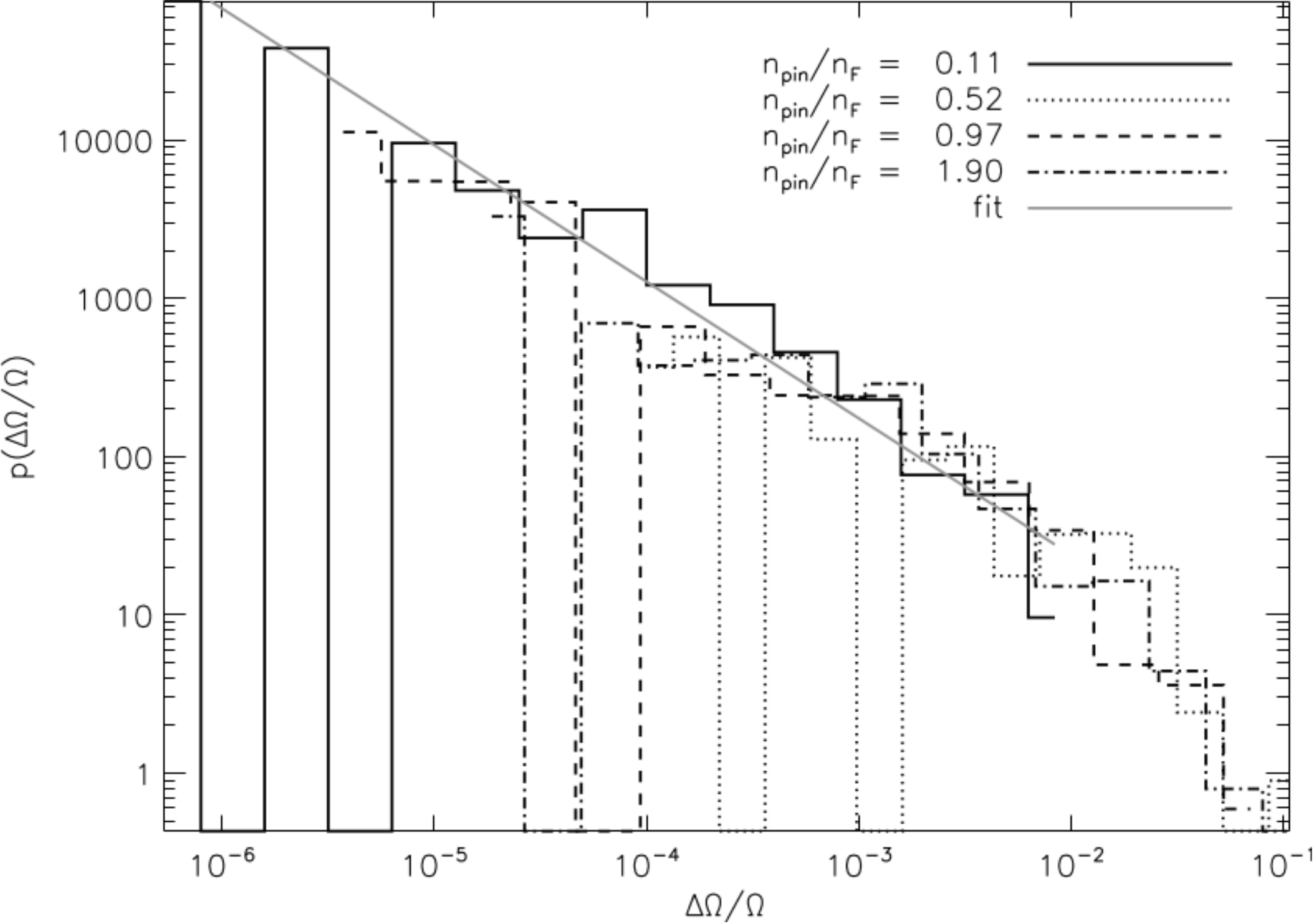}\;\;\;\;\;\includegraphics[width=8.5cm]{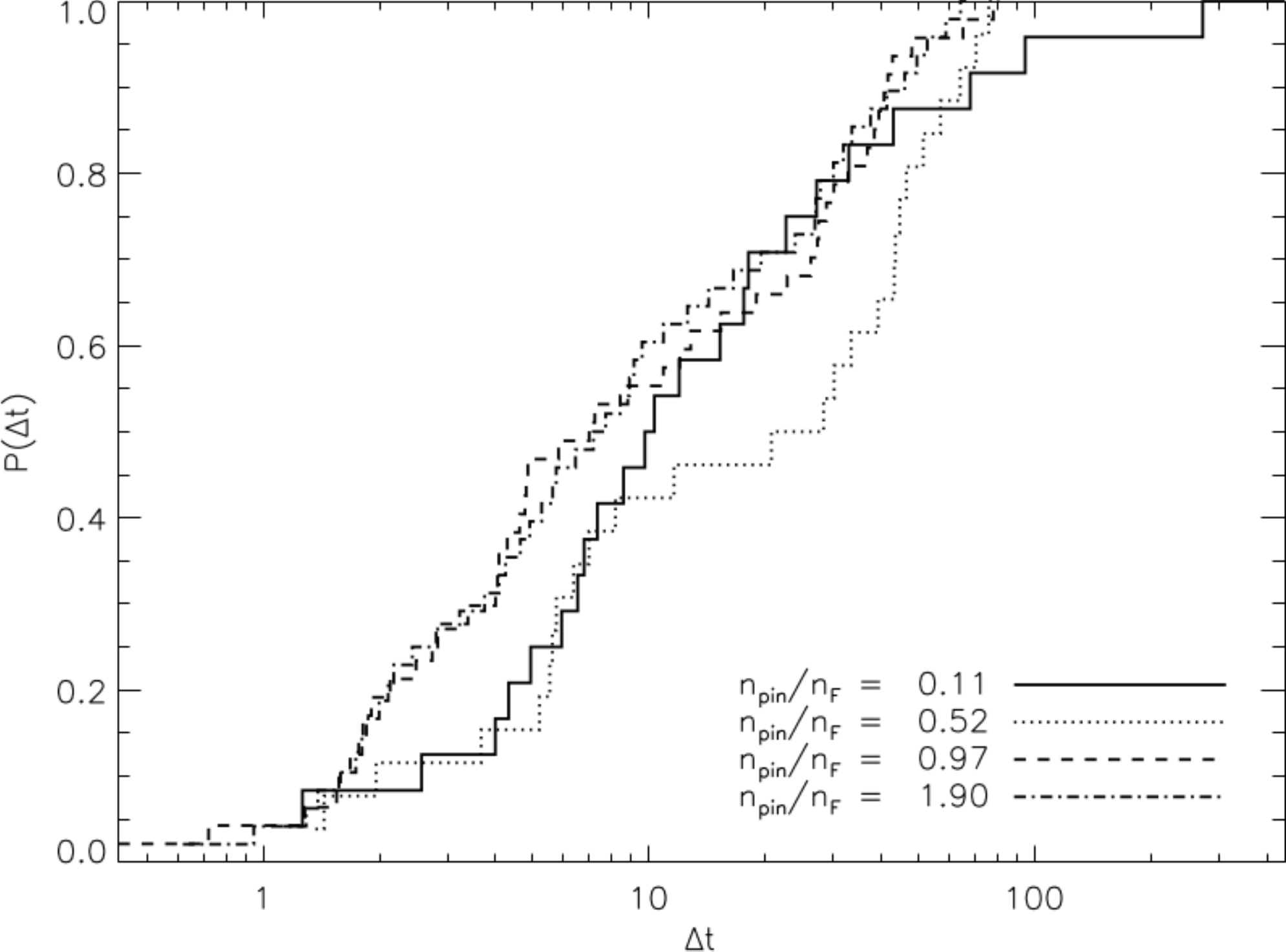}}
\vspace*{8pt}
\caption{Probability density function of fractional glitch sizes (left) and cumulative probability function for waiting times between glitches (right), obtained from the simulations of Warszawski \& Melatos (2011) \cite{Lila11} with varying ratio of occupied to free pinning sites, $n_{\mathrm{pin}}/n_\mathrm{F}$ shown in figure \ref{lila1}. The grey line on the left hand graph is a power law least square fit to the solid histogram, with exponent -0.801. }\label{pdfs}
\end{figure}

Warszawski and Melatos \cite{Lila11} carried out a series of numerical experiments with different parameters. In general, collective motion of vortices is observed, with vortex avalanches giving rise to glitches in the spin-down and distributions of sizes which obey power laws, with exponentially distributed waiting times. The main conclusions can be summarised as follows:
\begin{enumerate}
\item{ Vortices unpin in `herds' under the action of two knock-on processes: the proximity effect (unpinning due to vortex-vortex interactions) and acoustic knock-on (unpinning to sound waves shaking the vortex). The latter dominates the former when $\gamma$ is small and biases the dynamics towards the coherent noise model (Sec. 5.2), whereas the proximity effect dominates when $\gamma$ is large and the dynamics is more SOC-like (Sec. 5.3). This validates earlier suggestions by Alpar \cite{alpartrigger, capacitor} that random unpinning in capacitive vortex regions of the crust could initiate a vortex avalanche.}
\item{ Vortices tend to skip over pinning sites and more radially outwards by a few Feynman distances before knocking on a neighbour, thereby maintaining a roughly (but \emph{not} exactly) regular Abrikosov lattice. The pinning strength plays a fundamental role, with stronger pinning giving rise to larger glitches, and a broader range of potentials giving rise to a broader range of sizes and more frequent glitches \cite{Lila11}}
\item{A strong external spin-down torque leads to smaller, more frequent, glitches than a weak torque}
\item{A heavier crust produces more large glitches and longer waiting times than a lighter crust}
\item{Waiting times are longer and glitches larger when vortices and pinning sites are equally abundant, than if pinning sites are more abundant than vortices. Further increases in pinning site abundance do not change the results, as vortices tend to simply move across many pinning sites to readjust the Magnus force}
\end{enumerate}

In figure \ref{lila1} we show an example of different spin down curves for the crust obtained with different setups of the GPE code, with varying ratio of pinned to free vortices, $n_{\mathrm{pin}}/n_\mathrm{F}$, as described by Warszawski \& Melatos \cite{Lila11}. In figure \ref{pdfs} we also show the probability density function of fractional glitch sizes and the cumulative probability function for waiting times between glitches, obtained from these simulations. 

Motivated by the GPE results, Warszawski \& Melatos \cite{lilaknock} developed an asynchronous automaton (i.e. an automaton in which the time step is itself a stochastic variable that depends on the state of the system, see eg. \cite{Cornforth05} for a description of how this choice influences the system) to study the statistics of pulsar glitches. They show that a simple Poisson process in which vortices unpin independently (e.g. by thermal creep) cannot explain the broad distribution of glitch sizes observed in the pulsar population, even if one accounts for variable pinning strengths (with the only difference in this case being that stronger pinning sites are preferentially occupied, leading to occasional large glitches). In order to obtain a broad power-law size distribution knock-on effects need to be included (i.e. one has to allow for the possibility that upon unpinning a vortex will initiate an avalanche). Nevertheless also in this case a power law distribution is only obtained if the spin-down rate and unpinning rate (which can be expected to be a function of temperature if we allow mainly for thermal creep) are fine tuned. This fine tuning is unlikely in an astronomical setting and further developments are required to understand this issue.

Finally let us note that the main limitation in applying the results discussed above directly to an astrophysical setting is linked with the scale of the simulations. In general in all the examples above vortices are separated by a few pinning sites, and the overall superfluid flow is strongly influenced by motion of the vortices themselves (hence the importance of knock-on effects). In a realistic neutron star the inter-vortex separation is much larger than the average distance between pinning sites, with an average vortex encountering up to $\approx 10^{10}$ pinning sites between its equilibrium position and its nearest neighbour. Two things can then happen. On the one hand, an unpinned vortex may skip over $\approx 10^{10}$ pinning sites before repinning. This is not as unlikely as it sounds, as the systems constantly self-adjusts to keep the effective Magnus-modified pinning potential marginally stable and roughly uniform everywhere. Alternatively vortices could be too sparse for proximity effects to play an important role and lead to knock-on effects. In this case vortices will creep out gradually, as we describe in the following section. The exact nature of the motion of a realistic vortex in a realistic neutron star thus remains an open, and central, problem.

\section{Hydrodynamical instabilites}

An important ingredient of most of the models described above is a pinned superfluid in the crust or core which can store angular momentum by lagging behind the `normal' fluid and the crust. An important implication of this picture, which has received little attention until recently, is that the superfluid analogue of the two-stream instability can occur \cite{Prix03, prix04}, possibly triggering vortex unpinning and hence a glitch.

One mode of oscillation triggered this way is the r-mode, a toroidal oscillation where the restoring force is the Coriolis force. This particular class of inertial modes has already attracted much interest as it can be driven unstable by gravitational wave emission \cite{NA98, morsink}. Its velocity field is particularly simple in the two-fluid description \cite{Hasrmode}, even when one accounts for a lag \cite{Nilsglitch, hogg}. In particular we  study linear perturbations of the equations in (\ref{system1})-(\ref{system3}), with a rotational lag $\Delta=(\Omega_\mathrm{p}-\Omega_\mathrm{n})/\Omega_\mathrm{p}$ in the background, as would be the case if vortices are pinned ($\mathcal{B}=0$ and $\mathcal{B}^{'}=1$ for strong pinning).
One can solve for a velocity field of the form (where $\mathrm{x}=\mathrm{p,n}$ labels the fluid) \cite{hogg}:
\be
\delta v_\mathrm{x}^i=-\frac{im}{r^2\sin\theta}U^l_\mathrm{x} Y_l^m \hat{e}_\theta^i+\frac{1}{r^2\sin\theta} U^l_\mathrm{x}\partial_\theta Y_l^m\hat{e}^i_\phi,
\ee
where $U_\mathrm{x}^l(r)$ are the mode amplitudes and $Y_l^m(\theta,\phi)$ are the standard spherical harmonics. Consider the simple case of vanishing entrainment and a constant density star, for which the equation of state of the perturbations is incompressible, i.e. ($\nabla_i\delta v_\mathrm{x}^i=0$). One can find simple solutions for the $l=m$ case, for which $U_\mathrm{c}^l=A_\mathrm{x} r^{l+1}$ and the mode frequency $\sigma$ takes the form \cite{hogg}:
 \beq
 \sigma&=&-\frac{1}{(m+1)x_\mathrm{p}}[1-x_\mathrm{p}+\Delta \pm D^{1/2}],\\
 D&=&(1+x_\mathrm{p})^2+2\Delta\lbrace 1+x_\mathrm{p}[3-m(m+1)]\rbrace
 \eeq
 r-modes are therefore unstable for
 \be
 m > m_c=\frac{1+x_\mathrm{p}}{\sqrt{2x_\mathrm{p}\Delta}}.\label{mc}
 \ee
 The growth time for the instability is
 \be
 \tau_g\approx m \frac{P}{2\pi} \left(\frac{x_\mathrm{p}}{1+x_\mathrm{p}}\right)\left(\frac{m^2}{m_c^2}-1\right)^{-1/2},
 \ee
where $P$ is the rotation period of the star. 
The growth time scale must be compared to the time scale on which shear viscosity damps the mode, which, for a constant density star with $R=10$ km and $M=1.4 M_\odot$, is given by \cite{KS99}
\be
\tau_{sv}\approx \frac{6\times 10^{4}}{m^2} \left(\frac{T}{10^8 \mbox{K}}\right)^2 s;
\ee
that is, the mode grows provided that 
\be
m<500\left(\frac{\Delta}{10^{-4}}\right)^{1/4}\left(\frac{P}{0.01 \mbox{s}}\right)^{-1/2} \left(\frac{T}{10^8 \mbox{K}}\right).
\ee
This constraint, together with (\ref{mc}) tells us that the mode first becomes unstable at the critical lag
\be
\Delta_c\approx 6\times 10^{-5} \left(\frac{P}{0.1\mbox{ s}}\right)^{2/3}\left(\frac{T}{10^8 \mbox{K}}\right)^{-4/3}.\label{criticald}
\ee
Glampedakis and Andersson \cite{Nilsglitch} used angular momentum conservation to compare (\ref{criticald}) to maximum glitch sizes and find $\Delta_c\leq (I_\mathrm{p}/I_\mathrm{n}) (\Delta/\Omega_\mathrm{p})$, suggesting that this mechanism can set the maximum size of a glitch, and larger lags are prevented by the instability (which may, in fact, trigger the glitch).

\begin{SCfigure}
  \centering
  \caption{A comparison of the critical lag in (\ref{criticald}) against data for glitches of several pulsars, with a given spin period $P$. The red dots are systems where the temperature can be measured, while for the triangles the temperature is estimated from a simple modified URCA model to be $T\approx 3.3\times 10^8 (t_s/1\mbox{yr})^{-1/6}$ K \cite{Nilsglitch}. The unstable region is in gray and $m_c$ represents the critical $m$ for the onset of the instability. For higher $m$ there is a range of unstable inertial modes. A ratio $I_\mathrm{p}/I_\mathrm{n}\approx 0.02$ is assumed and the ratio $\Delta\Omega_\mathrm{c}/\Omega_\mathrm{c}$ used in Glampedakis and Andersson is equivalent to $\Delta/\Omega$.}\;\;\;\;\;\;
  {\includegraphics[width=10cm]{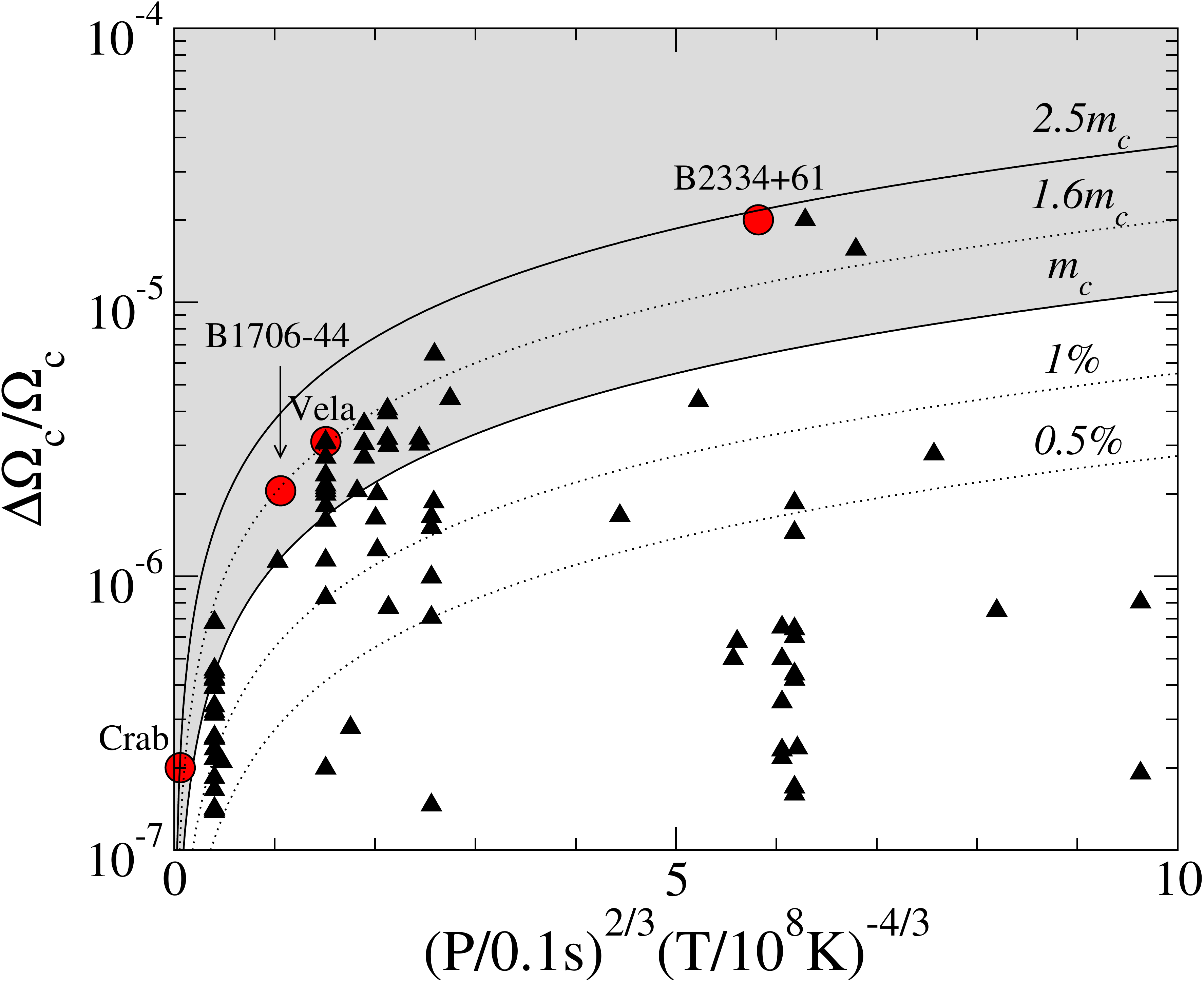}}\;\;\;\;\;\;\;\;\;\;
 \label{nils}
\end{SCfigure}
In figure (\ref{nils}) we show a comparison of the prediction for the critical lag with data, for an assumed value of the ratio $I_\mathrm{p}/I_\mathrm{n}\approx 0.02$. The model can explain the different maximum sizes in different pulsars by accounting for differences in temperature (which is inferred from standard cooling models \cite{Nilsglitch}) and age. A more complex analysis, including entrainment \cite{hogg}, reveals  that not only does the instability persist, but that a secular (long timescale) instability exists for low $l$ in the pinned regime.

Link has also shown, in a plane wave analysis, that instabilities still exist, both in the core and the crust, if one does not assume perfect pinning but allows vortices to creep out, i.e. a regime such that $\mathcal{B}<<\mathcal{B}^{'}$ \cite{Link12core, Link12Crust}. The instability is thus unlikely to be stabilised by viscosity or mutual friction and can play a significant role in glitches.

Another important instability in laboratory superfluids is the Donnelly-Glaberson instability \cite{Glaberson74}. In superfluid helium this signals the transition to turbulence. When counterflow along the rotation axis exceeds the threshold $w_{DG}=2\sqrt{2\nu\Omega}$, Kelvin waves destabilise the vortex array and create a vortex tangle. For standard parameters $|w_{DG}|\approx1.24\left({0.1\mbox{ s}}/{P}\right)^{1/2}$, where $\nu=(\kappa/4\pi)\log{b_0/a_0}$, with $b_0$ the inter vortex spacing and $a_0$ the radius of the vortex core \cite{Peraltaglitch}.

In a realistic neutron star the crust is spun down first by electromagnetic torques, and the rest of the fluid follows in response to magnetic torques or Ekman pumping. Peralta et al. \cite{Peralta05, Peraltaglitch} showed that, in the absence of magnetic fields, a sizeable axial counterflow develops which exceeds the threshold for the Donnelly-Glaberson instability in most of the star. The mutual friction force changes form and strength in those regions where the threshold is exceeded. Note that these calculations are carried out in the Hall-Vinen-Bekarevich-Khalatnikov (HVBK)  \cite{HVBK1, HVBK2} formalism that describes superfluid helium, i.e. a superfluid condensate and its massless excitations. Nevertheless the results should not depend strongly on this choice and highlight the possibility that turbulence is likely to play an important role in pulsar glitches, with transitions between laminar and turbulent mutual friction consistent with the short spin-up and long inter-glitch timescales observed \cite{Peraltaglitch}. Recently Melatos and Link \cite{ML14} have also shown that shear-driven turbulence exerts a fluctuating torque on the crust, which may contribute to the timing noise observed in pulsars.

Mastrano and Melatos \cite{MMglitch} suggested that a Kelvin-Helmholtz instability at the interface between the $^1S_0$ and $^3P_2$ superfluids in the star may trigger the glitch, complementing the bulk two-stream instability discussed above.

\section*{Relaxation}

\begin{SCfigure}
  \centering
  \caption{The inferred superfluidity coefficients $\mathcal{B}E^{-1/2}$ (where $E$ the Ekman number, i.e. $E=\eta/\rho_\mathrm{n}R^2\Omega^2$ with $\eta$ the shear viscosity coefficient) plotted against $\rho_\mathrm{n}(1+K)$, where $K$ is the ratio of the total moment of inertia of the fluid to that of the crust (See van Eysden \& Melatos \cite{vE10} for details). The lightly and dark shaded regions correspond to the range inferred from Vela and Crab data, respectively. The theoretical curves are calculated by van Eysden \& Melatos  \cite{vE10} for three different temperatures: $T=10^{7.5}$ K (top, red), $T=10^{7}$ K (centre, green) and $T=10^{6.5}$ K (bottom, blue). In these models $\rho_\mathrm{n}$ is varied for a fixed value of $K=50$.}\;\;\;\;\;
  {\includegraphics[width=9cm]{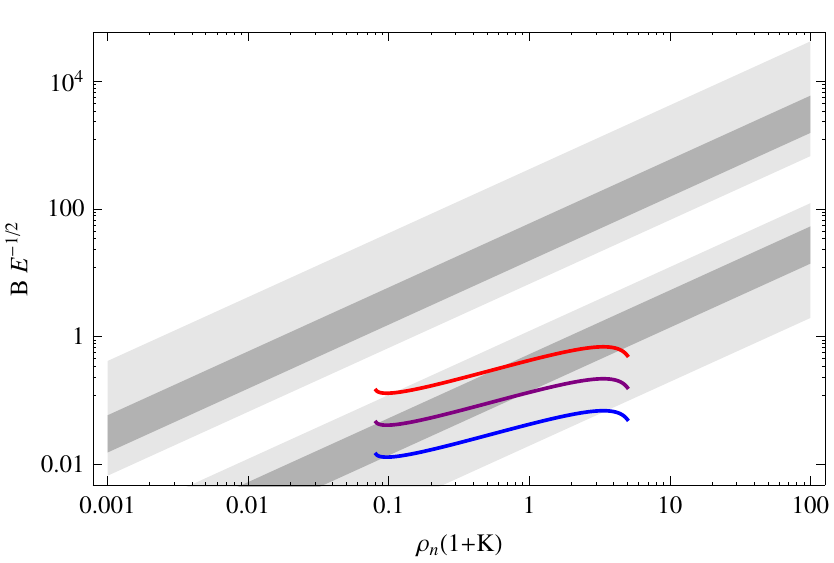}}\;\;\;\;
 \label{vE10fig}
\end{SCfigure}

The long timescales (weeks to months as described in section \ref{pheno}) observed in post-glitch relaxation were the first evidence that neutron stars contain a superfluid component, loosely coupled to the crust \cite{2fluid}. Two classes of models have been advanced. Alpar and collaborators \cite{Alpar84a, Alpar84b} suggested that, by analogy with flux creep in superconductors, vortices `creep' out as the star spins down, and the response of the creep rate to a glitch determines the spin rate of the star and gives rise to the observable relaxation. Alternatively, if pinning is weak in the crust, as suggested by Jones \cite{JonesNOPIN}, the vortices move with the neutron superfluid, and mutual friction is the dominant mechanism in the relaxation .
Van Eysden and Melatos \cite{vE10} analysed the general case where the ratio of the Ekman and mutual friction time-scales can be small or large. They obtained good agreement with Crab and Vela data as well as laboratory experiments with superlfuid helium (to an accuracy of 0.5 \% in the latter case \cite{vE12}).
From the analytic solution of  van Eysden and Melatos \cite{vE10} one can also obtain a general condition for observing an 'overshoot' in the relaxation, i.e. for observing the post-glitch frequency to drop below the value it would have had in the absence of a glitch and subsequent relaxation, as is the case for the Crab, but not Vela.
In the model of van Eysden and Melatos \cite{vE10} this is found to occur if immediately after the glitch one has $\Omega_\mathrm{p}<\Omega_\mathrm{n}<\Omega+\Delta\Omega_G$, i.e. after a glitch of size $\Delta\Omega_G$ the crust is spinning slower than the superfluid component. The different behaviour of the Crab and Vela pulsars would then suggest a difference in the internal vortex and/or pinning site distribution. The viscosity coefficients of nuclear matter obtained by fitting the relaxation of both pulsars are, however, consistent with each other, and with the theoretical predictions of van Eysden and Melatos \cite{vE10}, as can be seen from figure \ref{vE10fig}.

\section{Vortex creep}
\label{Creep}

In a seminal work Alpar et al. \cite{Alpar84a}, inspired by the analogy with flux creep in type II superconductors, developed a theory of non-linear coupling between the crust and the neutron fluid, due to vortex `creep' between pinning sites and applied it to the recovery of Vela giant glitches \cite{Alpar84b}.
The main idea is that the core superfluid couples rapidly to the normal component, due to electron scattering on vortex cores \cite{AlparMF1}, according to (\ref{mfcore}). The observed dynamics must then arise from the crust. This conclusion was supported by the observed increase in spin-down rate after the glitch, which points to a few percent of the moment of inertia being decoupled, the appropriate amount for the crust superfluid. Short term increases in the spin-down rate observed in later glitches, however, can exceed 100\%, implying that `decoupling' the superfluid in the crust is not enough; one needs a time-dependent internal torque (e.g. a change in mutual friction mechanisms)\cite{Link92, Miri05}.

Vortices in the crust pin to ions \cite{AI, AlparPin1}. On long time-scales, however, the neutrons follow the spin-down of the crust, as vortices migrate from one pinning site to the next, via thermal excitations or quantum tunnelling \cite{Link93}, i.e. they `creep' out. The Magnus force biases the motion in the outward direction, leading to an average radial velocity $\langle v_r \rangle \neq 0$. Suppose one views the Magnus force $f^i_{M}=\rho_\mathrm{n}\kappa^i \varpi \Delta\Omega$ as arising from a fictitious pinning energy \cite{Alpar84a}
\be
\Delta E_p=E_p\frac{\Delta\Omega}{\Delta\Omega_c},
\ee
where $\Delta\Omega_c$ is the critical lag for unpinning (i.e. after which the pinning force can no longer balance the Magnus force). For outward radial motion, the energy barrier that a vortex must overcome is lowered to
\be
E_+=E_p-\Delta E_p=E_p(1-\Delta\Omega/\Delta\Omega_c),
\ee
while for motion in  the $-r$ the barrier is higher:
\be
E_-=E_p+\Delta E_p=E_p(1+\Delta\Omega/\Delta\Omega_c).
\ee
In other words for a vortex to move radially {\it in} work has to be done against the sum of the pinning force and the Magnus force that tends to push the vortex out. For a vortex to move out, on the other hand, work has to be done against a lower total force, given by the pinning force minus the Magnus force.

In thermal equilibrium the radial bias in the pinning potential leads to more vortices randomly unpinning and moving out than in, and thus to an average outward creep speed:
\be
v_r=v_0\left[\exp\left(-\frac{E_+}{kT}\right)-\exp\left(-\frac{E_-}{kT}\right)\right]\approx v_o\exp \left(-\frac{E_+}{kT}\right),\label{vcreep}
\ee
where $v_0$ is the typical velocity of microscopic motion of the vortex lines between pinning centres, which Alpar et al. \cite{Alpar84a} take to be due to the Bernoulli force, with $v_0\approx 10^7$ cm/s. Note that this assumed value is critical to the calculation, and is simply taken to be a constant in the original model. Several authors have estimated the average speed of free vortices, and found it to vary significantly in different regions, and depart from the constant value given above \cite{JM06, HaskellLagrange, Link13}, leading to significant differences in the predictions of the theory, as we shall see in the following.  More detailed calculations of $E_{\pm}$ \cite{LE91} account also for the tension of the vortex line; vortices behave rigidly in weaker pinning regions, unpinning simultaneously from several sites, while they are more flexible in stronger pinning regions (in which the pinning force dominates over the tension), with unpinning dominated by single-site events. The relative importance of the tension can be expressed in terms of a `stiffness' parameter $\tau_s=\hat{T}r_0/F_p\bar{l}$, where $\hat{T}$ denotes the tension, $F_p$ is the maximum pinning force, $r_0$ is the range of the pinning potential and $\bar{l}$ is the internuclear spacing\cite{LE91}:
\be
\tau_s \approx 10 \left(\frac{\rho_\mathrm{n}}{10^{14}\mbox{g/cm$^3$}}\right)\left(\frac{E_p}{1\mbox{MeV}}\right)^{-1}\left(\frac{r_0}{10\mbox{fm}}\right)^{2}\left(\frac{\bar{l}}{50\mbox{fm}}\right)^{-1}.
\ee
In the limiting cases of $\Delta\Omega\approx \Delta\Omega_c$ and $\Delta\Omega<<\Delta\Omega_c$, the activation energy $E_a$ (not to be confused with the pinning energy per site $E_p$) to unpin can be written simply as:


\beq
E_a&\approx& 2.2 E_p \sqrt{\tau_s}\frac{\Delta\Omega_c}{\Delta\Omega}\;\;\;\;\mbox{for}\;\;\;\Delta\Omega<<\Delta\Omega_c,\\
E_a&\approx& 5.1 E_p \sqrt{\tau_s}\left(1-\frac{\Delta\Omega}{\Delta\Omega_c}\right)\;\;\;\;\mbox{for}\;\;\;\Delta\Omega\approx\Delta\Omega_c
\label{eactive}
\eeq


In order to obtain a radial creep rate Link et al. \cite{Link93} also estimate the rate at which vortices `attack' (either due to thermal excitations or to quantum mechanical zero point motion) the barrier and subsequently repin. They find that
\beq
v_r&\approx&\frac{\omega_s}{2\pi}l_{min} j_* \left(\frac{2\pi k T_{eff}}{E_a}\right)^{1/2} e^{-E^*_a/k T_{eff}},\;\;\;\;\;\mbox{for}\;\;\;\tau_s>1\label{refine1}\\
v_r&\approx&\frac{\omega_f}{2\pi}l_{min}e^{-E^*_a/k T_{eff}},\;\;\;\;\;\mbox{for}\;\;\;\tau_s<1\label{refine2}
\eeq
where the asterisks indicate that the quantities are evaluated for the characteristic number of pinning bonds $j_*$ involved in an unpinning event, $\omega_s$ and $\omega_f$ are the characteristic Kelvin oscillation frequencies for stiff and flexible vortices \cite{LE91}, $l_{min}$ is the minimum distance between pinning sites, and one has $T_{eff}=T_q \coth(T_q/T)$, where $T_q$ is the crossover temperature above which vortices surmount the barrier thermally, and below which quantum tunnelling through the barrier dominates.

In order to make quantitative comparisons with glitch recoveries Alpar et al. \cite{Alpar84b} assume that there exist physically distinct regions of different pinning strengths. Subsequent calculations do, in fact, show variations of pinning energies with density, and random variations of the pinning force due to different orientations of vortices \cite{DP, Avogadro, Seveso14}. Depending on the temperature, a given region with a given pinning strength, can be in either the linear or non-linear regime of vortex creep \cite{Alpar89}. If $E_p$ exceeds the binding energy of a nucleus to its equilibrium site, so that it is energetically favourable for a vortex line to dislodge nuclei, one has `strong' pinning. When this is not the case, but the vortex core radius $\xi$ (the superfluid coherence length) is smaller than the lattice spacing, one has `weak' pinning. Finally, when the superfluid gap decreases and the vortex core is large enough to encompass several nuclei, there is little change in energy between nearby configurations (as the number of pinned nuclei remains approximately constant as the vortex moves) and one has `super weak' pinning. 

The equations of motion for the system are
\beq
I_\mathrm{p}\dot{\Omega}_\mathrm{p}&=&N_{ext}+\sum_i I^i_\mathrm{n}\frac{2\Omega^i_{\mathrm{n}}}{\varpi}v_r,\label{creep}
\eeq
where $\Omega_p(t)$ is the angular velocity of the crust and all components tightly coupled to it, $\Omega^{i}_{n}(t)$ is the angular velocity of the superfluid in the $i^{th}$ pinning region, and
$v_r$ is obtained in the $i^{th}$ pinning regions from the expression in (\ref{vcreep}), with $\Delta\Omega^i=\Omega_p-\Omega^{i}_{n}$. The external spin-down torque is given by $N_{ext}$ and $I_\mathrm{n}$ is the moment of inertia of the crust and all components tightly coupled to it, which in the formulation of Alpar et al. \cite{Alpar84a} includes the entirety of the core neutrons, which are assumed to be coupled on timescales much shorter than those associated with the glitch \cite{AlparMF1}. The glitch trigger is not specified; it is introduced as an initial condition. Elsewhere Alpar et al. \cite{alpartrigger} discuss how random unpinning of vortices in `capacitive regions' (i.e. stronger pinning regions with an overdensity of vortices) can lead to vortex avalanches, a suggestion that has indeed been verified by GPE simulations \cite{lilaknock}.

Given an initial angular velocity perturbation $\Delta\Omega_i(0)$ induced by a glitch there are two regimes of dynamical response of the equation in (\ref{creep}). If the temperature is sufficiently large compared to the pinning energy, then a steady state (in which the neutrons spin down at the same rate as the crust) can be achieved with a small enough lag that the response of the spin-down rate to the glitch is linear in the size of the perturbation. The contribution of the $i_{th}$ pinning region shows up in the post-glitch response $\Delta\dot{\Omega}_\mathrm{p}$ as an exponential decay:
\be
\Delta\dot{\Omega}_{\mathrm{p},i}(t)=-\frac{I_i}{I}\frac{\Delta\Omega_i(0) e^{-t/\tau_{l,i}}}{\tau_{l,i}},
\ee
where $I_i$ is the moment of inertia of the region, $I$ is the total moment of inertia of the star, and the local coupling timescale in the $i^{th}$ is given by:
\be
\tau_{l,i}=\frac{kT}{E^i_p}\frac{\Delta\Omega^i_c \varpi}{4\Omega v_o} e^{E^i_p/kT}.
\ee

If, on the other hand, the temperature is low compared to the pinning energy, the steady state  lag is large and the response to the perturbation is non linear. In this regime the effect of a region $k$ on the response of the spin down rate has a Fermi function dependence on the glitch size, of the form:
\be
\frac{\Delta\dot{\Omega}_{\mathrm{p},k}}{\dot{\Omega_{\mathrm{p}}}}(t)=\frac{I_k}{I}\left(1-\frac{1}{1+[\exp(t_{o,k}/\tau_{n,k})-1]\exp(-t/\tau_{n,k})}\right),
\ee
where $I_k$ is the moment of inertia of the region, $t_{0,k}=\Delta\Omega(0)/|\dot{\Omega}_\mathrm{n}|$ and the relaxation time can be expressed as:
\be
\tau_n=\frac{kT}{E_p}\frac{\Delta\Omega_{c}}{|\dot{\Omega}_\mathrm{n}|}.
\ee
Creep islinear if $\tau_l<\tau_n$, with the transition occurring at a pinning energy of $E_p/kT\approx 31$ \cite{Alpar89}, corresponding to timescales $\tau\approx 1000$ days for Vela, for weak pinning. Given that this is roughly the periodicity of Vela glitches it is clear that non-linear relaxation will be difficult to observe. However, as we shall discuss in the following section, the timescale is much shorter if the transition occurs in a super weak pinning zone, and potentially observable. In fact most of the timescales involved in the relaxation of Vela glitches are interpreted as linear response of different regions, with only the longest timescales open to the possibility of a non-linear response (although combinations of linear responses are possible) \cite{Alpar93}. The observation of a Fermi-function like behaviour would, on the other hand,  suggest the presence of non-linear creep in the star.

To explain the observations of the Vela pulsar Alpar and collaborators develop a 'minimalist' glitch model \cite{Alpar84a, Alpar89, Alpar93}. They assume that vortices accumulate in a 'capacitive' region A1, which is adjacent to a depletion region B, in which no vortex motion is present (i.e. is not contributing to the long-term spin down). During the glitch vortices unpin from region A1, pass through region B and finally repin in a region further out, A2. These are the 'active' regions, while there are 'passive' regions further in the star in which no vortex motion occurs and that react to the glitch only via the change in lag induced by the glitch (i.e. for these regions $\Delta\Omega=\Delta\Omega_\mathrm{glitch}$).

After integrating the contributions for these different regions, one obtains the general form for the post glitch response as \cite{Alpar93}:
\beq
&& \frac{\Delta\dot{\Omega}_\mathrm{p}}{\dot{\Omega}_\mathrm{p}}=\left(\frac{I_A}{I}+\frac{I_B}{I}\right)\Theta(t)-\frac{I_A}{I}\frac{t}{t_g}\Theta(t)\Theta(t_g-t)-\left(\frac{I_A}{I}+\frac{I_B}{I}\right)\Theta(t-t_g)+\nonumber\\
&&+\int \frac{d I_{l}}{I}\frac{\Delta\Omega_{l}(0)e^{-t/\tau_l}}{|\dot{\Omega}_\mathrm{p}\tau_l}+\int\frac{dI_{nl}}{I}\left(1-\frac{1}{1+[\exp(t_{o}(t)/\tau_{n}-1]\exp(-t/\tau_{n})}\right),\label{relax}
\eeq
where the subscript $l$ refers to regions in the linear coupling regime and $nl$ to regions in the nonlinear coupling regime. $I_A$ is the moment of inertia of region A1 (the region from which vortices depin, $I_B$ of the depletion region through which vortices travel, while $I_{A2}<<I_{A1}$ and is neglected. One has $t_g=\Delta\Omega(0)/|\dot{\Omega}_\mathrm{n}|$ for regions A and B.

By fitting the expression in (\ref{relax}) to the post-glitch relaxation of eight Vela glitches, Alpar et al. \cite{Alpar93} reach the following general conclusions:
\begin{itemize}
\item{Short timescale exponential relaxation generally corresponds to the response of passive regions  (i.e. regions through which no vortex motion occurs)  in the linear regime, which are identified with super weak pinning regions with pinning energies of around $E_p\approx 0.3$ MeV.}
\item{The full non-linear Fermi function response is not observed in the recoveries of Vela glitches}
\item{The remaining longer timescales and linear terms in the relaxation are more difficult to interpret, given that for $t_0<<\tau$ the nonlinear term reduce to an exponential. If one has relaxation timescales $\tau$ of approximately a month and glitches of approximately $\Delta\Omega_\mathrm{p}\approx 10^{-4}$ rad/s, then $t_0\approx 10 \mbox{days} <<\tau$ and thus we are in the regime in which the nonlinear response will give rise to an exponential. One could then be observing the linear recoupling of two different regions, one with $\tau\approx 30$ days, and the other with $\tau>1000$ days. Such a long timescale would lead to what appears to be a permanent offset on shorter timescales. However Alpar and collaborators \cite{Alpar93} suggest that in reality one is observing non-linear recouping in super weak pinning, with regions through which no vortex motion has occurred giving rise to the $30$ day exponential relaxation and regions through which vortex motion occurred giving rise to the permanent increase in spin down rate. This places the unpinning regions close to the transition between linear and non-linear response, and the transition between super weak and weak pinning at slightly lower densities.}
\item{Alpar et al. \cite{capacitor, Alpar94} also suggest that the different glitching activity of the Crab is due to the star not having yet formed a fully developed network of capacitor regions, and that crust quakes may be the trigger for unpinning for this pulsar (as a opposed to vortex avalanches for the Vela pulsar).}
\end{itemize}

In general the theory is successful in describing the relaxation of Vela glitches with consistent parameters. Although there are some difficulties in explaining large increases in the spin down rate on short timescales of a minute or so after the 2000 and 2004 Vela glitches \cite{JM05}, we will see in the following that these can be overcome if one allows for part of the core to decouple after a glitch. This is also necessary, as described above, if one is to explain the activity of the Vela pulsar while accounting for strong entrainment in the crust. Another important point is that most of the recovery can be described by exponentials, i.e. by creep in the linear regime. In this regime creep can be approximated in the two-fluid picture in terms of a fraction of free vortices \cite{JM06}, allowing us to apply the formalism described in section (\ref{hydro}).
Nevertheless the presence of several free parameters to account for different pinning strength and coupling timescales leads to a theory that is, in fact, difficult to falsify. Recently Link \cite{Link13} has examined the problem of vortex creep in more detail His conclusion is that both the linear regime and super weak pinning invoked by standard vortex creep theory described above are unlikely to be realised in practice. Let us review Link's revised theory of vortex creep, dubbed vortex 'slippage'.

\subsection{Vortex slippage}

Recently Link \cite{Link13}has revised the vortex creep model, by generalising two of the main simplifying assumptions of the original theory. The first is the form of the pinning barrier, for which Link \cite{Link13} uses more realistic values, such as those in equations (\ref{eactive}), the second, is the typical velocity of an unpinned vortex, which in the original theory is fixed at $v_0\approx 10^7$ cm/s. In vortex slippage theory the velocity of a free vortex moving from one pinning site to the next is obtained by accounting for tension and drag forces, resulting in an average velocity for a vortex of the form \cite{Link13}:
\be
<v>=r\Delta\Omega \left (\frac{1}{2}\sin 2\theta_d \hat{r}+\cos^2\theta_d \hat{\phi}\right)e^{-(E_a/kT)}
\ee
where $\theta_d$ is the dissipation angle, related to the mutual friction coefficient by $2\mathcal{B}=\sin 2\theta_d e^{-E_a/kT}$.
The post glitch response is found to depend non-linearly on the size of the glitch, with pinned regions responding on a time-scale
\be
t_d\gtrsim7 \left(\frac{t_{sd}}{10^4\mbox{yr}}\right)\left(\frac{\Delta\Omega_G/\Omega}{10^{-6}}\right)\mbox{  days},\label{delay}
\ee
where $t_{sd}$ is the spin down timescale and $\Delta\Omega_G$ the size of the glitch. Exponential decay can only occur on timescales longer than $t_d$, and thus the shorter exponential timescales observed in the recovery of Vela glitches are unlikely to be due to vortex slippage in either the crust or core. Haskell et al. \cite{Haskell12} interpret the latter timescales as due to free vortices coupled by mutual friction in the outer core (see section \ref{MFsec}).
Finally Link estimates  the pinning strength, also accounting phenomenologically for strong entrainment, and the possibility of pinning both to the crustal lattice and to flux tubes in the core. His calculations, together with the estimates for the coupling timescale in (\ref{delay}) suggest that the linear creep regime introduced by Alpar and collaborators \cite{Alpar84a, Alpar89, Alpar93} is not realised in practice and that super-weak pinning is is precluded by thermal fluctuations. 
Observations of a delayed response in the recovery after a glitch, over a timescale proportional to the glitch size as in (\ref{delay}) would, on the other hand, would provide a hint that vortex slippage is occurring in neutron stars.

Finally let us note that vortex creep is a  dissipative process and creep heating \cite{Alpar84a, LLreheat, SL89, Reisenegger} might explain the unusually high (for their age) temperatures that have been measured for some old, recycled millisecond pulsars, such as J0437-4715 \cite{UV} .

\subsection{Thermal glitches}

Temperature clearly plays an important role in the creep model. In the previous discussion however the trigger mechanism for the glitch itself is left unspecified and one does not account for heat deposition during the glitch and how this would affect the motion of vortices. This problem was examined by Link and Epstein \cite{LE96} who considered the effect of depositing heat in a region where the glitch is triggered (e.g. due to the strain released by a quake, or dissipation due to mutual friction during a vortex avalanche). If the trigger is a crust quake, one could deposit up to $10^{42}(\bar{\sigma}_{c}/10^{-2})$ erg in the crust \cite{LE96}. After the trigger deposits heat locally, a heat front travels through the crust, increasing the creep rate and spining-up of the crust further. The glitch then ends either when the thermal energy diffuses or when the superfluid and crust velocities are equal. This scenario was further considered by Larson and Link \cite{LarsLink} who simulated Crab and Vela glitches by solving the equations of motion for creep in (\ref{creep}), with the detailed prescription for the creep velocity in (\ref{refine1}-\ref{refine2}), coupled to the thermal diffusion equation
\be
c_v\frac{\partial T}{\partial t}=\nabla_i (\kappa_T\nabla^i T) + h_f-\epsilon_\nu,\label{heat}
\ee
where $c_v$ is the specific heat, $\kappa_T$ the thermal conductivity, $\epsilon_\nu$ the neutrino emissivity and $h_f$ the heating rate per unit volume due to superfluid friction in the crust, which corresponds simply to the rotational energy dissipated in the crust due to vortex motion \cite{Alpar84a, vR95} and is obtained from $ H=\int dI_\mathrm{n}\Delta\Omega|\dot{\Omega}_\mathrm{n}|$, dividing by the volume of the crust. The simulations are spherically symmetric and initiated by depositing heat (corresponding to typical energies of $10^{42}$ erg) in a spherical shell centred on a density of $\rho=1.5\times 10^{14}$ g/cm$^4$ with a width of 40 m.

The results of the simulations are compared to glitches in several pulsars, e.g. Crab and Vela. The agreement is good, with the interesting conclusion that, for similar heat depositions, glitches in hotter pulsars (such as Crab) are generally smaller and slower than in older pulsars (such as Vela), essentially due to the larger specific heat \cite{LarsLink}. This model thus accounts for the continued rise observed in the 1989 Crab glitch for a day (but is not commonly observed), which the simulations of Larson and Link do not reproduce if the glitch is introduced as a sudden jump in frequency with no heat deposition. Furthermore the large heat deposition can lead to thermal pulses at the surface, which are close to the current observational upper limits on post-glitch heating of Vela (see section \ref{pheno}).

\section{Mutual friction}
\label{MFsec}
The creep paradigm relies on vortices being pinned in the crust, and only a small number 'creeping' out. Clearly if the pinning is very weak and most vortices are free this will not be the case. This was suggested by Jones \cite{JonesNOPIN}, who pointed out that for a infinitely long vortex the pinning interactions from the surrounding ions would average out, leading to it being essentially free. He asuggested that this also happens for a realistic, rigid vortex, and that vortices in the crust are close to co-rotation with the neutron superfluid. The observed recovery timescales are then due to mutual friction. Although recent calculations reveal that, for realistic tension forces, there is still a sizeable pinning force \cite{Seveso14}, the fact that creep is expected to be in the linear regime for most of the post-glitch relaxation lets one simulate a glitch using of the multi-fluid formalism described in section (\ref{hydro}).
Haskell et al. \cite{Haskell12} have integrated the equations in (\ref{system1}-\ref{system3}) in cylindrical symmetry by averaging the mutual friction coefficients over the length of a vortex. Mutual friction is taken to be due to electron scattering in the core, with a mutual friction coefficient of the order $\mathcal{B}\approx 10^{-4}$ given by (\ref{mfcore}), and to phonons in the crust, with a mutual friction coefficient of the order $\mathcal{B}\approx 10^{-9}$ \cite{KelvonJones, PhononJones}. 


\begin{figure}
\centerline{\includegraphics[width=8.6cm]{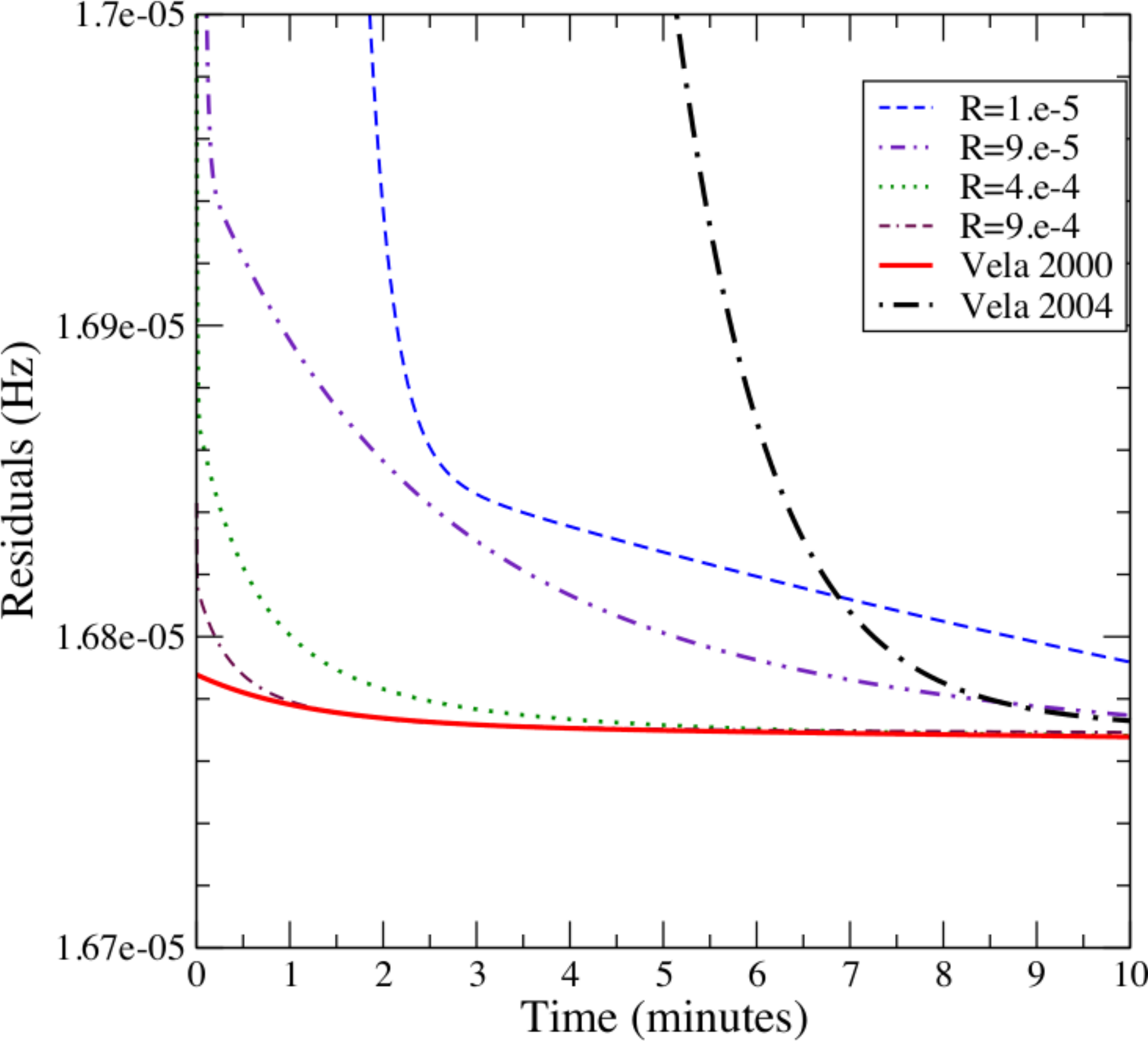}\;\;\;\;\;\;\;\includegraphics[width=8.5cm]{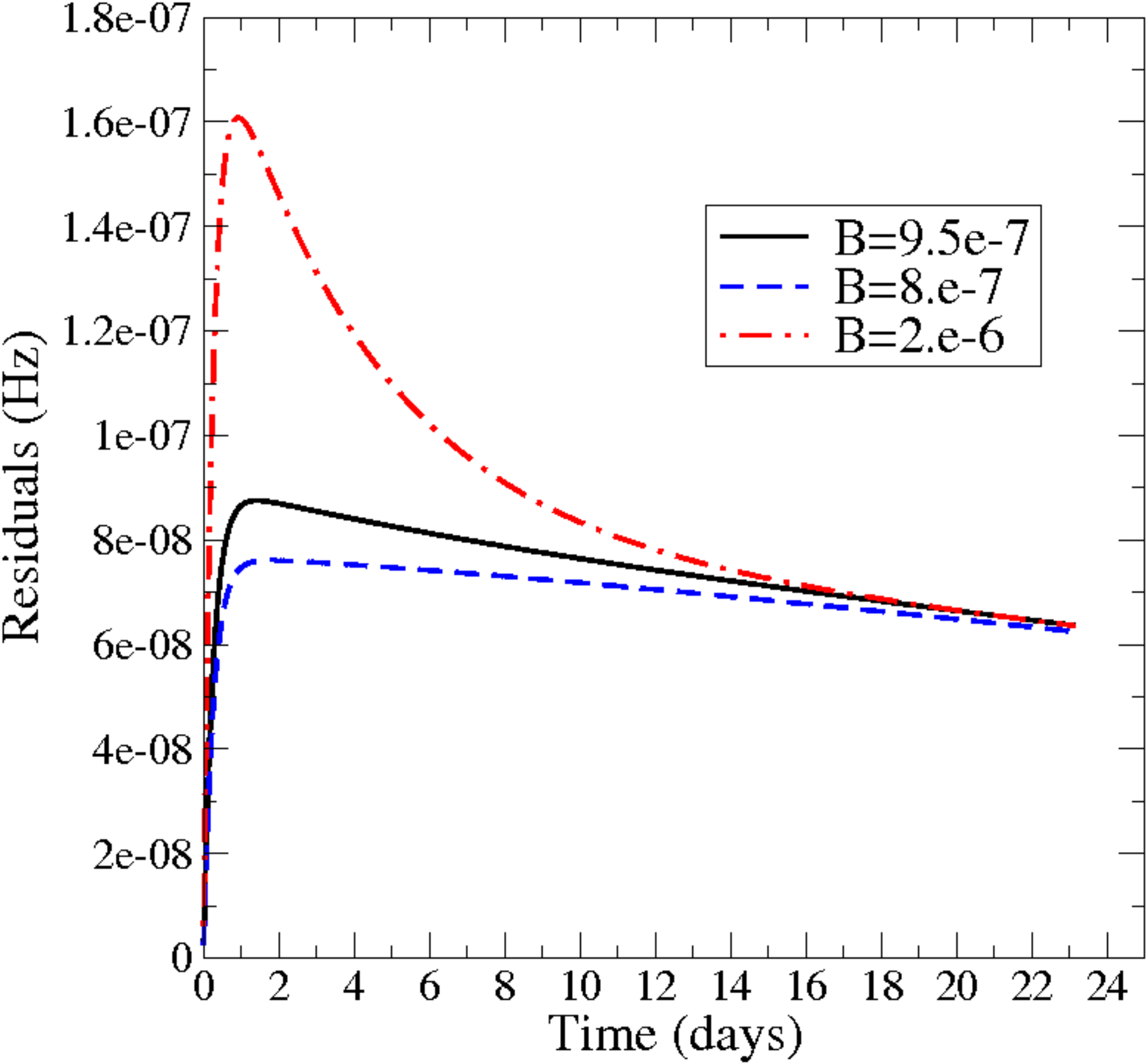}}
\vspace*{8pt}
\caption{LEFT: Comparison of the short-timescale post glitch relaxation simulated by Haskell et al. \cite{Haskell12} for varying values of the drag parameter $\mathcal{R}$ in the core,  to the observed relaxation of the Vela 2000 and 2004 glitches. The pre glitch spin down has been subtracted. In general we can see that larger values of $\mathcal{R}$ are favoured, with $\mathcal{R}\approx 10^{-4}$, as expected from electron scattering of vortex cores, consistent with the data. Values $\mathcal{R}<10^{-5}$, as would be the case if a large number of vortices remain pinned to flux tubes in the core, are not consistent with the data, given this model.
RIGHT: Example, in the setup of Haskell and Antonopoulou \cite{HA14} of the different relaxation behaviours one can have by increasing $\tilde{\mathcal{B}}$ during the glitch, in an 80 m region of the crust, with a background (pre-glitch) $\mathcal{B}=5\times 10^{-9}$. As $\tilde{\mathcal{B}}$ increases (and the rise time decreases) $\Delta\nu/\nu$ passes from 0.9$\%$ for $\mathcal{B}=8\times 10^{-7}$, to 1.3$\%$ for $\mathcal{B}=9.5\times 10^{-7}$, and finally for $\mathcal{B}=2\times 10^{-6}$ an exponential trend begins to appear.}\label{relax}
\end{figure}

\begin{figure}
\centerline{\includegraphics[width=14cm]{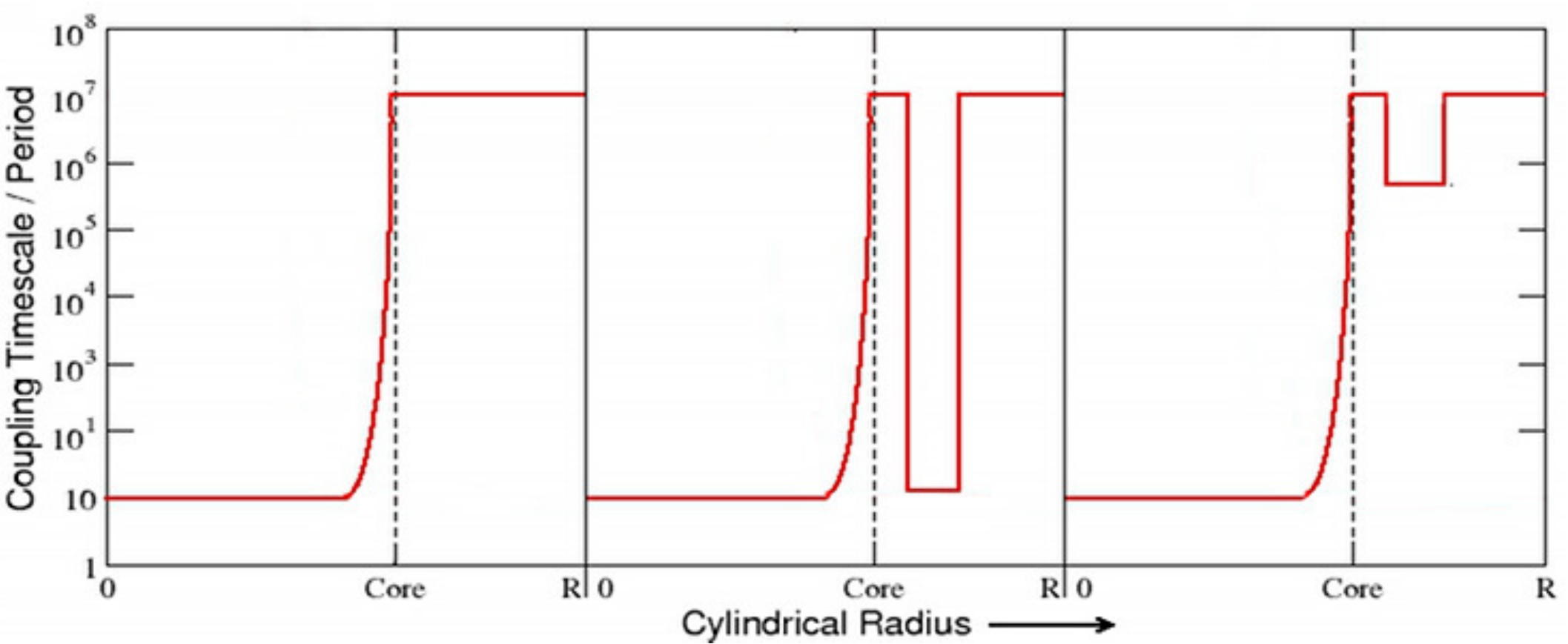}}
\vspace*{8pt}
\caption{The left panel shows a schematic representation of the coupling in the crust and outer core. In the middle panel we show the large decrease in coupling timescale that would lead to a glitch that decouples part of the core, and then relaxes (such as a giant glitch in the Vela). In the right panel we show the situation for an avalanche that frees only a small number of vortices and thus only involves the crust. In the picture of Haskell and Antonopoulou \cite{HA14} this gives rise to glitches for which the relaxation appears essentially as a step in frequency and frequency derivative, with no exponential relaxation.}\label{dec}
\end{figure}

The system is evolved with both fluids originally co-rotating and all vortices in the crust pinned, i.e. $\mathcal{B}=0$ (note that Haskell et al. \cite{Haskell12} ignore the effect of entrainment, which reduces the lag accumulated during spin down). As the evolution proceeds the lag $\Delta\Omega$ is compared to a realistic pinning profile (such as those described in section \ref{snowplow}) and if it exceeds the critical lag for unpinning $\Delta\Omega_c$ mutual friction is switched back on, mimicking the effect of vortices unpinning. The region then relaxes back to equilibrium and no re-pinning is allowed until after a glitch. Once the lag reaches the maximum of the critical lag (we remind the reader that the critical lag varies with density, and thus has a maximum in the crust, as described in section \ref{snowplow}), a glitch is initiated. This is done by assuming that as vortices move out they excite Kelvin waves, thus leading to strong dissipation and rapid recoupling. In practice this is achieved by switching the mutual friction parameter to $\mathcal{B}_k\approx 10^{-3}$ in the region near the maximum that hat has not yet relaxed to its equilibrium lag, i.e. for which $\Delta\Omega>\dot{\Omega}_\mathrm{p}/2\Omega_\mathrm{n} \mathcal{B}$, where $\mathcal{B}$ is the crustal mutual friction parameter before the switch to $\mathcal{B}_k$.
This leads to a rapid rise (typically of the order of seconds) and glitches of the order of $\Delta\Omega/\Omega\approx 10^{-6}-10^{-5}$. 

In figure \ref{relax} we show fits to the Vela 2000 and 2004 glitch recoveries. In general the results are consistent with electron scattering in the core as the main source of mutual friction, but not with the drag being much weaker, as could be the case if vortices pin to flux tubes in the core during the glitch.

Haskell and Antonopoulou \cite{HA14}  adapted the above model to smaller glitches, to show that random unpinning will lead not only to glitches of different sizes, but also to glitches with different rise times. This follows from introducing a fraction of unpinned vortices $\gamma$ in the equations of motion ({\ref{system1}-\ref{system3}) as an effective mutual friction coefficient $\tilde{\mathcal{B}}=\gamma\mathcal{B}$. The coupling timescale between the fluids is thus a function of $\gamma$ and is given by $\tau\approx 1/(2\Omega\tilde{\mathcal{B}})$. Large values of $\gamma$ (i.e the unpinning of many vortices) produce preferentially larger glitches that decouple part of the core, as shown in figure \ref{dec} (and as in the previously described simulations of large Vela glitches), and are followed by exponential relaxations. Small values of $\gamma$ lead to glitches that do not involve the core and look like simple steps in frequency and frequency derivative, as can be seen in figure \ref{relax}. This picture is also applied to relaxations of magnetar glitches, for which large increases in the spin-down rate can be obtained on long timescales simply because the magnetars are much slower than ordinary pulsars, leading to the recouping timescales for the interior regions $\tau\approx 1/(2\Omega\mathcal{B})$, which give rise to the recovery, being consequently increased.

An effect that is ignored in the above picture is that of Ekman pumping, which will act to communicate changes in the spin of the crust to the bulk fluid of the interior. van Eysden and Melatos showed that the timescales can be comparable to the shortest in glitch relaxations in the case of non-magnetized stars \cite{vE10}, but much shorter in the presence of magnetic fields \cite{vE14}. In the non-magnetised case the problem can be solved exactly in different geometries for varying ratios of the Mutual Friction and Ekman timescale\cite{vEJLT}. However, in the magnetic case, the dynamical response of the fluid is complex, exhibiting various modes of oscillation after a spin up (down) event, which may be observable in post-glitch relaxations \cite{vE14}.

\section{Gravitational Waves}
Gravitational radiation is neither scattered nor absorbed. It is therefore an ideal probe of dynamical processes in neutron star interiors, which often lack direct electromagnetic signatures. The next generation of ground-based, long-baseline, gravitational-wave interferometers are on schedule to commence science operations in the next few years \cite{Riles}, with sensitivites $\sim$10 times greater than previously achieved. It is therefore timely to review briefly the possibilities for multi-messenger (e.g. simultaneous radio and gravitational-wave) studies of pulsar glitches in the coming Advanced Detector Era.

Gravitational wave signals from a glitch divide into two types: broadband bursts from the spin-up event itself, and narrowband continuous-wave emission from the inter-glitch interval and post-glitch recovery. The burst signal is triggered by whatever non-axisymmetric dislocation triggers the spin-up event, e.g. crust cracking or superfluid vortex unpinning. The energy released can excite the acoustic and intertial stellar oscillations, e.g. f and p modes \cite{K01, Sed03, Keer} and their superfluid versions \cite{Sidery10}, generating an oscillating mass quadrupole moment lasting $\lesssim$ 0.1 s. The energy conversion efficiency is poorly known \cite{AC01, SP14}. Alternatively, the rapid motion of the unpinned vortices, before they repin, can produce a time-varying current quadrupole moment, again on the spin-up time-scale \cite{LilaGW}. Current quadrupole radiation is also emitted by glitch-triggered r-modes \cite{SP12}. The continuous-wave signal is generated by residual non-axisymmetries in the superfluid velocity field, some parts of which are erased on the post-glitch recovery time-scale. Flow non-axisymmetries arise from Ekman circulation \cite{vE08, Bennett10} and/or oscillatory vortex structures at high Reynolds number, e.g. Taylor-Gortler modes \cite{Peraltaglitch, MP10}. The emission is through the current quadrupole channel. However, if glitches arise from vortex avalanches, then some part of the above non-axisymmetries persist beyond the recovery stage and the pinned vortex distribution following a glitch is non uniform, as in other SOC system \cite{Jensen98}. A general argument based on angular momentum conservation in the context of glitches triggered by vortex avalanches shows that the persistent, current-quadrupole wave strain ${h}_0$ from a glitching pulsar satisfies \cite{Mprep}
\be
h_0 \gtrsim 3 \times 10^{-21} (\Omega/10^3\text{ rad s}^{-1})^3 ({d}/1\text{kpc})^{-1} ({I_\mathrm{p}/I_\mathrm{n}}) (\Delta\Omega_{max}/\Omega),
\ee
where $d$ is the distance to the source, and $\Delta\Omega_{max}/\Omega$ is the maximum fractional glitch size over the current spin-down time-scale, whether or not that largest glitch occurred during the gravitational wave observation.

At the time of writing, the most sensitive searches for gravitational radiation from glitches have been performed with Science Run 5 data from the Laser Interferometer Gravitational wave Observatory (LIGO). A Bayesian analysis of the August 2006 glitch in the Vela pulsar, incorporating prior information about known instrumental artefacts, placed a 90\% confidence upper limit of ${h}_0 \leq 1.4\times10^{-20}$ on the wave strain amplitude of quasinormal mode ring-down signals with frequency 1-3 kHz and decay time 0.05-0.5 s \cite{Clark07, LIGOglitch}. The same search, conducted with Advanced LIGO, would be $\sim$ 10 times more sensitive. Advanced detector searches for post-glitch, continuous-wave signals are also planned using algorithms for long-duration transients with phase drifts, e.g. Bayesian window template \cite{Prix11} and frequency-time cross-power \cite{Thrane11}. If the quadrupole is generated by pinned vortices, it co-rotates with the crust, hence is phase locked to the radio ephemeris \cite{Mprep}.

Finally, looking even further into the future beyond the Advanced Detection Era, instruments like the Einstein Telescope are being designed with enhanced Newtonian noise suppression in the 10-50 Hz band, where many glitching pulsars reside. The prospects for further enlargement of the glitch discovery space are therefore bright.


\bibliographystyle{ws-ijmpd}
\bibliography{glitchrev}
\end{document}